\documentclass[twocolumn,floatfix,tighten]{aastex62}

\received{}
\revised{}
\accepted{}
\submitjournal{ApJ}

\usepackage{graphicx}	
\usepackage{amsmath}	
\usepackage{amssymb}	
\usepackage{bm}		
\usepackage{graphbox}
\usepackage{booktabs}
\usepackage{listings}
\usepackage{color}
\usepackage{tablefootnote}
\usepackage{enumerate}

\usepackage{blkarray, bigstrut}
\usepackage{lineno}

\usepackage{appendix}
\usepackage{xr}
\definecolor{dkgreen}{rgb}{0,0.6,0}
\definecolor{gray}{rgb}{0.5,0.5,0.5}
\definecolor{mauve}{rgb}{0.58,0,0.82}

\newcommand{\DAP}{{\tt DAP\,}}

\let\oldAA\AA
\renewcommand{\AA}{\text{\normalfont\oldAA}}

\usepackage[T1]{fontenc}
\usepackage{ae,aecompl}

\usepackage{newtxtext,newtxmath}

\usepackage{xcolor}
\definecolor{alizarin}{rgb}{0.82, 0.1, 0.26}

\shortauthors{Schaefer et al.}
\shorttitle{Local scaling relations for N/O}
\begin{document}

\title{SDSS-IV MaNGA: Exploring the local scaling relations for N/O}

\author{Adam L. Schaefer}
\altaffiliation{schaefer@mpa-garching.mpg.de}
\affiliation{Max-Planck-Institut f{\"u}r Astrophysik, Karl-Schwarzschild-Str. 1, D-85748 Garching, Germany}
\affiliation{Department of Astronomy, University of Wisconsin-Madison, 475N. Charter St., Madison, WI 53703, USA}

\author{Christy Tremonti}
\affiliation{Department of Astronomy, University of Wisconsin-Madison, 475N. Charter St., Madison, WI 53703, USA}

\author{Guinevere Kauffmann}
\affiliation{Max-Planck-Institut f{\"u}r Astrophysik, Karl-Schwarzschild-Str. 1, D-85748 Garching, Germany}

\author{Brett H. Andrews}
\affiliation{University of Pittsburgh, PITT PACC, Department of Physics and Astronomy, Pittsburgh, PA 15260, USA}

\author{Matthew A. Bershady}
\affiliation{Department of Astronomy, University of Wisconsin-Madison, 475N. Charter St., Madison, WI 53703, USA}
\affiliation{South African Astronomical Observatory, P.O. Box 9, Observatory 7935, Cape Town, South Africa}
\affiliation{Department of Astronomy, University of Cape Town, Private Bag X3, Rondebosch 7701, South Africa}

\author{Nicholas F. Boardman}
\affiliation{Department of Physics and Astronomy, University of Utah, 115 S. 1400 E., Salt Lake City, UT 84112, USA}

\author{Kevin Bundy}
\affiliation{UCO/Lick Observatory, University of California, Santa Cruz, 1156 High St. Santa Cruz, CA 95064, USA}

\author{Niv Drory}
\affiliation{McDonald Observatory, The University of Texas at Austin, 1 University Station, Austin, TX 78712, USA}

\author{Jos{\'e} G. Fern{\'a}ndez-Trincado}
\affiliation{Instituto de Astronom\'ia, Universidad Cat\'olica del Norte, Av. Angamos 0610, Antofagasta, Chile}

\author{Holly P. Preece}
\affiliation{Max-Planck-Institut f{\"u}r Astrophysik, Karl-Schwarzschild-Str. 1, D-85748 Garching, Germany}

\author{Rog{\'e}rio Riffel}
\affiliation{Departamento de Astronomia, Instituto de F\'\i sica,
Universidade Federal do Rio Grande do Sul, CP 15051, 91501-970, Porto
Alegre, RS, Brazil}
\affiliation{Laborat\'orio Interinstitucional de e-Astronomia - LIneA, Rua
Gal. Jos\'e Cristino 77, Rio de Janeiro, RJ - 20921-400, Brazil}

\author{Rogemar A. Riffel}
\affiliation{Departamento de F\'isica, CCNE, Universidade Federal de Santa Maria, 97105-900, Santa Maria, RS, Brazil}
\affiliation{Laborat\'orio Interinstitucional de e-Astronomia - LIneA, Rua Gal. Jos\'e Cristino 77, Rio de Janeiro, RJ - 20921-400, Brazil}

\author{Sebasti{\'a}n F. S{\'a}nchez}
\affiliation{Instituto de Astronom{\'i}a, Universidad Nacional Aut{\'o}noma de M{\'e}xico, A.P. 70-264, 04510, Mexico, D.F., M{\'e}xico}

\begin{abstract}

We present, for the first time, the relationship between local stellar mass surface density, $\mathrm{\Sigma_{*}}$, and N/O derived from SDSS-IV MaNGA data, using a sample of $792765$ high signal-to-noise ratio star-forming spaxels. Using a combination of phenomenological modelling and partial correlation analysis, we find that $\mathrm{\Sigma_{*}}$ alone is insufficient to predict the N/O in MaNGA spaxels, and that there is an additional dependence on the local star formation rate surface density, $\mathrm{\Sigma_{SFR}}$. This effect is a factor of $3$ stronger than the dependence of 12+log(O/H) on $\mathrm{\Sigma_{SFR}}$. Surprisingly, we find that the local N/O scaling relations also depend on the total galaxy stellar mass at fixed $\Sigma_{*}$ as well as the galaxy size at fixed stellar mass. We find that more compact galaxies are more nitrogen rich, even when $\mathrm{\Sigma_{*}}$ and $\mathrm{\Sigma_{SFR}}$ are controlled for. We show that $\sim50\%$ of the variance of N/O is explained by the total stellar mass and size. Thus, the evolution of nitrogen in galaxies is set by more than just local effects and does not simply track the build up of oxygen in galaxies. The precise form of the N/O-O/H relation is therefore sensitive to the sample of galaxies from which it is derived. This result casts doubt on the universal applicability of nitrogen-based strong-line metallicity indicators derived in the local universe.


\end{abstract}

\section{Introduction}
Metallicities in both the stellar and gaseous components of galaxies have been an important tool for our understanding of galaxy evolution \citep[e.g.][]{Tinsley1980,Lilly2013}.

Recent years have seen a burgeoning in the volume of spatially resolved spectroscopic data from large surveys such as CALIFA \citep{Sanchez2012}, SAMI \citep{Croom2012} and MaNGA \citep{Bundy2015}. These data have stimulated interest in the relationship between the gas-phase oxygen abundances, stellar mass, star formation rate and gas content of galaxies on kpc scales
\citep[e.g.][]{BarreraBallesteros2016,BarreraBallesteros2018,Mingozzi2020,Teklu2020,Wang2020}. By analogy to the global mass metallicity relation \citep{Tremonti2004}, the oxygen abundance has been shown to be sensitive to the local stellar mass surface density ($\mathrm{\mathrm{\Sigma_{*}}}$), tracing the integrated star formation history on local scales in a galaxy, as well as the presence of outflows \citep[e.g][]{BarreraBallesteros2018} and inflows \citep{Lian2019,Schaefer2019}. Since early observations of this local relationship between $\mathrm{\Sigma_{*}}$ and $12+$log(O/H) \citep{Moran2012,RosalesOrtega2012}, there has been a growing consensus that the global mass-metallicity relation can be explained as arising on local scales. That is, the accumulation of chemical elements on kpc scales can be seen as a reflection of the buildup of stellar mass locally within galaxies over their evolutionary history rather than on global scales \citep[see e.g.][]{Sanchez2020ARAA,Sanchez2021}. This is reinforced by recent observations that gas-phase chemical abundances in galaxies are correlated on $\sim \mathrm{kpc}$ scales \citep{Sanchez2015,Kreckel2020,Li2021} implying that their chemical enrichment  proceeds by the local injection and diffusion of metals into the ISM of galaxies. Nevertheless, some observations have shown that the local gas-phase metallicity is also related to the total stellar mass of galaxies \citep{Gao2018}. This is likely due to the greater depth of the gravitational potential well of more massive galaxies, which makes the expulsion of metals through feedback-driven outflows more difficult.

\cite{Mannucci2010} found that the form of the global mass metallicity relation varied with the star formation rates of galaxies, such that at fixed stellar mass, the oxygen abundance is lower for galaxies with greater star formation rates. This is explained by galaxy chemical evolution models as the result of the accretion of low-metallicity gas simultaneously diluting the interstellar medium (ISM) of the galaxy and triggering an increase in the star formation rate \citep[see e.g.][]{Lilly2013}. The existence of this so-called `Fundamental Mass-Metallicity Relation' (FMR) on local scales within galaxies is still not confirmed. \cite{Teklu2020} find that the local FMR is present in a sample of MaNGA galaxies for the N2H$\alpha$ and O3N2 abundance indicators \citep{Pettini2004}, but is not seen with N2O2 \citep{Kewley2002} or N2S2H$\alpha$ \citep{Dopita2016}.

Some part of the inconsistency of the FMR between different studies may be due to the sensitivity of some strong-line oxygen abundance indicators to N/O \citep{Kashino2016}. The N/O abundance ratio does not respond to changes in a galaxy's evolutionary state in the same way as O/H. As such, studies of N/O are sensitive to different evolutionary processes in galaxies than O/H. For example, a tight scaling between gas-phase N/O and the integrated stellar mass of galaxies was observed by \cite{PerezMontero2013}. This was not observed to vary strongly with the star formation rate as was the case with O/H. The weak dependence of N/O on SFR at fixed stellar mass was likewise observed by \cite{Andrews2013}, this time using the direct method to determine chemical abundances. \cite{Kashino2016} used the N2S2H$\alpha$ oxygen abundance diagnostic, which estimates O/H via a correlation with N/O, and similarly found no SFR-dependence for the Mass-N/O Relation (the `Fundamental Mass N/O Relation'; FMNOR). \cite{Kashino2016} suggested that the lack of a secondary dependence of N/O on SFR is easily explained by the accretion of pristine gas. The additional fuel that enhances the SFR should dilute N and O by the same amount, leaving N/O unchanged. This effect has the potential to explain why the FMR is not seen with O/H indicators based on N2O2 and N2S2H$\alpha$ in other studies. However, the assumption that gas is accreted with pristine abundances is not valid in the low-redshift universe, with a substantial fraction of it having been already enriched by feedback-driven outflows \citep{Oppenheimer2010,Peng2014}. The observed lack of a SFR-dependence in the FMNOR is accompanied by a redshift evolution that is slow in comparison to O/H, leading some authors to suggest that it can be used as a fundamental probe of galaxy evolution \citep[e.g.][]{PerezMontero2013,Masters2016}. N/O is therefore an abundance ratio that should be investigated and fully understood.


The numerical chemical evolutionary modelling of \cite{Vincenzo2016} showed that the processes determining the relative abundances of nitrogen and oxygen are complex and cannot be explained by the effects of dilution alone. This complexity stems from the different mechanisms by which nitrogen and oxygen are released into the interstellar medium. The majority of oxygen is produced in massive stars through the $\alpha$-process and then released into the ISM by Type II supernovae within $\sim 10$ Myr of the onset of an episode of star formation \citep{B2FH,Leitherer1999}. However, the fraction of the total nitrogen budget of a galaxy produced in massive stars (called Primary nitrogen) is small, and dominates only in galaxies with low metallicity ($\mathrm{Z\lesssim 0.2\, Z_{\odot}}$). At higher metallicity, a significant fraction of nitrogen is produced by the CNO cycle in low and intermediate mass stars and then dispersed into the ISM in the final stages of stellar evolution. The yield of nitrogen in this case depends on the initial amounts of carbon and oxygen, leading to the correlation between N/O and O/H. The disparate timescales for the production of oxygen and nitrogen in galaxies is the origin of the complexity inherent in modelling the N/O ratio.

In their models, which considered galaxies as single objects with no substructure, \cite{Vincenzo2016} showed that the N/O at a given O/H is influenced by several factors. These include the rate at which gas is being accreted (infall timescale), the rate at which it is being consumed (the star formation efficiency; SFE), the ratio of massive stars to low-mass stars (the stellar initial mass function; IMF), and the relative rates at which these elements are ejected from galaxies by winds (the outflow loading factors). Comparing their models to SDSS single-fibre spectroscopic data, they concluded that some combination of these effects must be invoked to explain the observed N/O - O/H relation, but that no model was able to match the observed ratios without different outflow loading factors for O and N in the winds.

\cite{Matthee2018} used the EAGLE hydrodynamical simulations \citep{Crain2015,Schaye2015,McAlpine2016} to explore the relationship between N/O and SFR per unit mass. They find that the delayed production of nitrogen following a starburst leads naturally to a correlation between specific SFR (sSFR) and N/O at a given stellar mass. Given that their simulations indicate that the sSFR is a good indicator of the star formation history of a galaxy, the interpretation of this result is that the N/O ratio is also sensitive to the integrated star-formation history.

The relationship between N/O and O/H when considering galaxies in a resolved sense becomes more complicated still. Within galaxies the mobility of these elements through galactic fountain flows \citep{Shapiro1976} and the radial migration of stars \citep[e.g.][]{Elbadry2016} cannot be ignored, and the spatial variation of the SFE  \citep{Leroy2008} and IMF \citep{Parikh2018} may also play a role. Indeed, \cite{Belfiore2017} showed that the N/O - O/H relation varies systematically with the total stellar mass of galaxies such that the relation is flatter in more massive systems. Further exploration of this phenomenon by \cite{Schaefer2020} showed that differences in the SFE are a plausible explanation for some (but not all) of the variation in N/O at fixed O/H. With so many factors predicted to influence relative abundance of nitrogen and oxygen, it is surprising that N/O and stellar mass, or N/O and O/H correlate as tightly as they do.

In this paper, we will study the relationship between N/O, O/H and various local and global properties of galaxies to determine how these scaling relations are set. For the first time, we will investigate how N/O scales locally with $\mathrm{\Sigma_{*}}$, and the star formation rate surface density, $\mathrm{\Sigma_{SFR}}$. Understanding the nitrogen abundance in galaxies will allow us to trace star formation on a different timescale to oxygen, providing a unique view into the chemical evolution of galaxies.

The layout of this paper is as follows: In Section \ref{Methods} we summarise the spectroscopic data used for our study, the data selection criteria and the measurements made on the data to derive out conclusions. Section \ref{Results} contains the main results of our analysis, which we discuss in detail in Section \ref{Discussion}. We present our conclusions in Section \ref{Conclusions}.

All measurements assume a standard $\Lambda$ Cold Dark Matter cosmology, with $\Omega_{m}=0.3$, $\Omega_{\Lambda}=0.7$ and $H_{0}=70 \, \mathrm{km s^{-1} \, Mpc^{-1}}$. Unless stated otherwise, all stellar mass and star formation rate estimates assume a \cite{Chabrier2003} stellar IMF.

\section{Methods}\label{Methods}
\subsection{The data}
This study makes use of data obtained with the SDSS-IV MaNGA Survey. MaNGA is a large integral field spectroscopic survey that was performed on the $2.5$ m SDSS telescope at Apache Point Observatory \citep{Gunn2006} as part of the 4th stage of the SDSS endeavour \citep{Blanton2017}. The final MaNGA survey has accumulated data for approximately $10000$ galaxies \citep{Wake2017}. The $17$ optical fibre hexabundles range in size from $19$ to $127$ $2\arcsec$ fibres, covering a hexagonal region of sky between $12\arcsec$ and $32\arcsec$. The light collected by the hexabundles is passed to the BOSS spectrograph \citep{Smee2013}, where it is dispersed with a spectral resolution of $R=\lambda/\Delta\lambda\approx 2000$ \citep{Law2021} and covering a broad range of wavelengths between $3600$ and $10300\, \mathrm{\AA}$. The resulting spectra are allocated to a square grid of $0\farcs5\times0\farcs5$ spaxels by the MaNGA data reduction pipeline \citep{Law2016}, and smoothed to a spatial resolution of $2\farcs5$. The data reduction pipeline provides spectra that are calibrated to approximately percent-level accuracy \citep{Yan2016b}.
For more information on the MaNGA instrument, observing strategy and survey design, see \cite{Drory2015}, \cite{Law2015} and \cite{Yan2016}. 
Our results are based on the measurement of emission line fluxes measured from the MaNGA data cubes. There have been a number of independent efforts to measure the fluxes. We will make use of the fluxes derived by the MaNGA Data Analysis Pipeline \citep[\DAP;][]{Westfall2019}. These are derived by fitting and subtracting the stellar continuum fitted using Penalized Pixel Fitting \citep[pPXF;][]{Cappellari2004,Cappellari2017}. This method approximates the continuum with a linear combination of template stellar spectra from the MaStar stellar library \citep{Yan2019}. The emission line fluxes are approximated by a series of Gaussians, which are fitted simultaneously with the continuum after the stellar kinematics have been constrained. Fluxes derived in this way are robust and agree well with other non-parametric methods of line strength measurements \citep{Belfiore2019}.

To ensure that our conclusions are drawn from a clean sample of spectra we apply two independent sets of selection criteria to the data: One to select an appropriate sample of galaxies, and another to ensure the quality of the individual spaxel measurements.

\subsection{Galaxy Selection}\label{Galaxy_selection}
For this study we will use spaxel data from the MaNGA MPL-10 internal data release, which includes $9456$ unique galaxies. The sample selection criteria are very similar to those applied to \cite{Schaefer2019} and \cite{Schaefer2020}, though is drawn from a larger input sample. To ensure the robustness of our spaxel measurements we require face-on (Petrossian b/a>0.6), star forming galaxies. Within each galaxy, we determine the number of spaxels that show line emission attributable to star-formation. To do so, we select spaxels with observed [NII]/H$\alpha$ and [OIII]/H$\beta$ line ratios that satisfy the \cite{Kauffmann2003} and \cite{Kewley2001} criteria. We include the additional criterion that spaxels have H$\alpha$ equivalent widths of greater than $3 \, \AA$ in emission, to eliminate spaxels dominated by diffuse ionised gas \citep{CidFernandes2010}. If the fraction of spaxels that meet these conditions is below 0.6 (i.e. less than 60 per cent of the galaxy has usable data), then it is rejected. We include an additional constraint on the $r$-band Petrossian effective radii whereby galaxies are only included in our analysis if $R_{e}>4 \arcsec$. This criterion is included to minimise the impact of the $2\farcs 5$ MaNGA point spread function (PSF) on the measurement of local quantities in the presence of strong light gradients in our target galaxies. If a galaxy satisfies these criteria, then it is retained. We perform a thorough analysis of the impact of the PSF on our results in Appendix \ref{PSF_effect} and show that our main conclusions are unaffected by the spatial resolution of the MaNGA instrument. 
Finally, a visual inspection of the SDSS optical imaging of our sample yields two galaxies for which foreground stars have disrupted the measurements of the photometric structural parameters such as the $r$-band effective radius. We eliminate these galaxies from our sample. These criteria yield a total of $1497$ galaxies for analysis.
\subsection{Spaxel Selection}
Once the selection of galaxies has been made, we perform a number of measurements on the emission line fluxes from individual spaxels. To ensure the reasonable quality and reliability of our measurements, we only analyse spaxels that satisfy the following conditions. We require a minimum S/N ratio of $5$ in H$\alpha$, H$\beta$, [NII]$\lambda 6584$, [OIII]$\lambda 5007$, [OII]$\lambda 3726,29$ and [SII]$\lambda 6917,31$. Spaxels with the $\mathrm{H\alpha}/\mathrm{H\beta}<2.86$ will give an unphysical dust attenuation correction, so these are eliminated as well. 
Each spatial resolution element of MaNGA will incorporate emission from a veriety of sources. \cite{Lacerda2018} showed that diffuse ionised gas will contaminate H\textsc{ii} region spectra where the H$\alpha$ EW $<14 \, \mathrm{\AA}$, but \cite{ValeAsari2019} argue that a minimum $10 \, \mathrm{\AA}$ EW constraint on H$\alpha$ is sufficient to ensure that it does not dominate the measurements. In many cases, this can leave spectra with emission line ratios that are consistent with star formation, but which yield erroneous metallicity estimates. To reduce the impact of contamination from diffuse ionised gas, we opt for the slightly more relaxed requirement that the equivalent width of H$\alpha$ be above $10 \AA$ in emission. Our results do not change significantly with a more stringent constraint on the emission line equivalent widths, but the sample size is reduced. For spaxels that survive the H$\alpha$ EW and S/N cuts, we also reject those that have emission line ratios that are inconsistent with excitation from a young stellar population. For this purpose we use the \cite{Kauffmann2003} and the \cite{Kewley2001} criteria on the [NII]/H$\alpha$ - [OIII]/H$\beta$ ionisation diagnostic diagram. The application of the above criteria to our initial dataset yields a final sample of $792765$ spaxels for analysis. While the spaxel selection criteria are more stringent than those applied during the galaxy selection stage, the final spaxel sample does not eliminate any galaxies from our analysis.

\subsection{Metallicity Measurements}
MaNGA provides resolved spectra covering the entire optical band as well as the near infrared ($3600 \AA < \lambda < 10300 \AA$). This large spectral range provides the ability to estimate N/O and O/H using a variety of strong-line methods. While this paper will make use of several strong line estimators of both of these abundance ratios, the main results will use the estimator of \cite{Kobulnicky2004}, which is based on a combination of the R23 and O32 line ratios, where 
\begin{equation}
    \mathrm{R23 = \frac{[OII]\lambda 3726,29 + [OIII] \lambda 4959,5007}{H\beta}},
\end{equation}\label{R23_eq}
and
\begin{equation}
    \mathrm{O32=\frac{[OIII] \lambda 4959,5007}{[OII]\lambda3726,29}}.
\end{equation} \label{O32_eq}
This indicator uses the temperature-sensitive R23 ratio to constrain the overall oxygen abundance, while variations in the ionization parameter are taken into account by the O32 ratio. Using their prescription, the oxygen abundance can be written as
\begin{equation}\label{kk04_eq}
    \begin{aligned}
        12+\log(\mathrm{O/H})={}&9.11-0.218x-0.0587x^2 -0.330x^{3}\\ 
        &- 0.199x^{4}-y(0.00235-0.01105x\\
        &-0.051x^2 -0.04085x^3-0.003585x^4),
    \end{aligned}
\end{equation} 
where $x=\log(\mathrm{R23})$ and $y=\log(\mathrm{O32})$. This estimate of the oxygen abundance is valid only for $12+\log(\mathrm{O/H})>8.4$, but we note that the $99.64\%$ of spaxels in our dataset meet this condition.

To estimate the N/O ratio, we utilise the prescription of \cite{Thurston1996}. This method is based on the five-level atom calculation of \cite{Pagel1992}, and uses the ratio $\mathrm{[NII]\lambda 6548,84/[OII]\lambda 3726,29}$ as well as a small temperature-dependent correction based on R23,
\begin{equation}\label{NO_thurston}
    \log\left(\mathrm{\frac{N}{O}}\right)=\log\left(\frac{\mathrm{[NII]}}{\mathrm{[OII]}}\right)+0.307-0.02t_{\mathrm{II}}-\frac{0.726}{t_{\mathrm{II}}}.
\end{equation}
In Equation \ref{NO_thurston} $t_{\mathrm{II}}$, the temperature of the singly ionised oxygen zone of the H{\sc ii} region, is input in units of $10^{4} \, \mathrm{K}$.
The value of $t_{\mathrm{II}}$ in K is given by 
\begin{equation}
    t_{\mathrm{II}}=6065 + 1600x + 1878x^{2}+ 2803x^{3},
\end{equation}
where again, $x=\log(\mathrm{R23})$. \cite{Thurston1996} verified the accuracy of their calibration against a set of theoretical models and found that the absolute difference between their derived values and the modelled values was less than $0.1$ dex.

The chosen methods for deriving N/O and O/H are both based on theoretical models. They were chosen for our analysis because they appear to be the least systematically biased under variation in the ionisation parameter. For completeness, we reproduce some of the key Figures in Appendix \ref{Different_indicators} using both theoretical and empirically calibrated abundance estimators. These reproductions show that the main results of this paper are qualitatively robust to changes in the choice of abundance indicator, and that the conclusions of our work do not change if different strong-line estimators are used.

Our abundance estimates hinge on the measurement of the sums and ratios of different emission lines that are often separated in wavelength. To make these estimates as accurate as possible we correct for the effect of dust along the line of sight by comparing the observed Balmer Decrement ($f(\mathrm{H\alpha})/f(\mathrm{H\beta})$) to the Case B value of 2.86. Assuming that the dust is in the geometry of a foreground screen, we estimate the reddening using 
\begin{equation}
    E(B-V)=\frac{2.5}{k(\lambda_{\mathrm{H\beta}})-k(\lambda_{\mathrm{H\alpha}})}\log\frac{f(\mathrm{H\alpha})/f(\mathrm{H\beta})}{2.86},
\end{equation}
where $k(\lambda_{\mathrm{H\beta}})=3.66$ and $k(\lambda_{\mathrm{H\alpha}})=2.52$ are the values of the \cite{ODonnell1994} reddening curve assuming the $V$-band ratio of total to selective extinction, $R_{V}=3.1$. With this estimate of the reddening, we calculate the intrinsic flux of a line with wavelength $\lambda$ to be
\begin{equation}
    F(\lambda)=f(\lambda)10^{0.4k(\lambda)E(B-V)}.
\end{equation}

\subsection{Star formation rate surface density}
To estimate the current rate of buildup of the stellar mass within a spaxel of a galaxy, we measure the star formation rate in a spaxel using the luminosity of H$\alpha$ emission \citep{Kennicutt1998}. We calculate the luminosity using 
\begin{equation}
    L(\mathrm{H\alpha})=F(\mathrm{H\alpha})4\pi d_{L}^{2},
\end{equation}
where $d_{L}$ is the luminosity distance inferred from the systemic redshift of the galaxy. The luminosity is converted to a star formation rate using 
\begin{equation}
    SFR=\frac{L(\mathrm{H\alpha})}{2.16\times10^{34} \, \mathrm{W}},
\end{equation}
which we then convert to a surface density by dividing by the projected area of the spaxel, correcting for inclination using the galaxy's $r$-band elliptical Petrossian $b/a$.

\subsection{Stellar mass surface density}
We utilise the stellar mass surface density estimates made available through the Pipe3D \citep{Sanchez2016a,Sanchez2016b,Sanchez2018} value added catalogue for MPL-10. The Pipe3D software bins contiguous spectra to a signal-to-noise ratio of $50$. These are then fitted with a linear combination of SSP models that cover four metallicities in the range $\mathrm{Z/Z_{\odot}}=0.2-1.5$ and fourteen ages between $1$ Myr and $14.5$ Gyr \citep{CidFernandes2013}. These fits give a mass-to-light ratio which is then scaled to the amount of light in each individual spaxel in the spatial bin to derive the stellar mass in each spaxel. The Pipe3D catalogues provide stellar masses assuming a \cite{Salpeter1955} stellar IMF. We correct the Pipe3D stellar mass surface densities to a \cite{Chabrier2003} IMF by multiplying by a constant factor of $0.62/1.06$ following \cite{Speagle2014}. To compute $\Sigma_{*}$, we divide this stellar mass by the area of each spaxel and make a correction for inclination by dividing the projected spaxel area by the $r$-band elliptical Petrossian b/a.

\section{Results}\label{Results}
Previous results using single-fibre spectroscopy \citep[e.g.][]{PerezMontero2013,Kashino2016} reported that N/O is tightly correlated with the integrated stellar mass of galaxies, with no secondary dependence on the star formation rate. In this section we will investigate whether these observations hold on local scales within galaxies.

\subsection{The scaling of N/O with local stellar mass surface-density}
\begin{figure}
    \centering
    \includegraphics{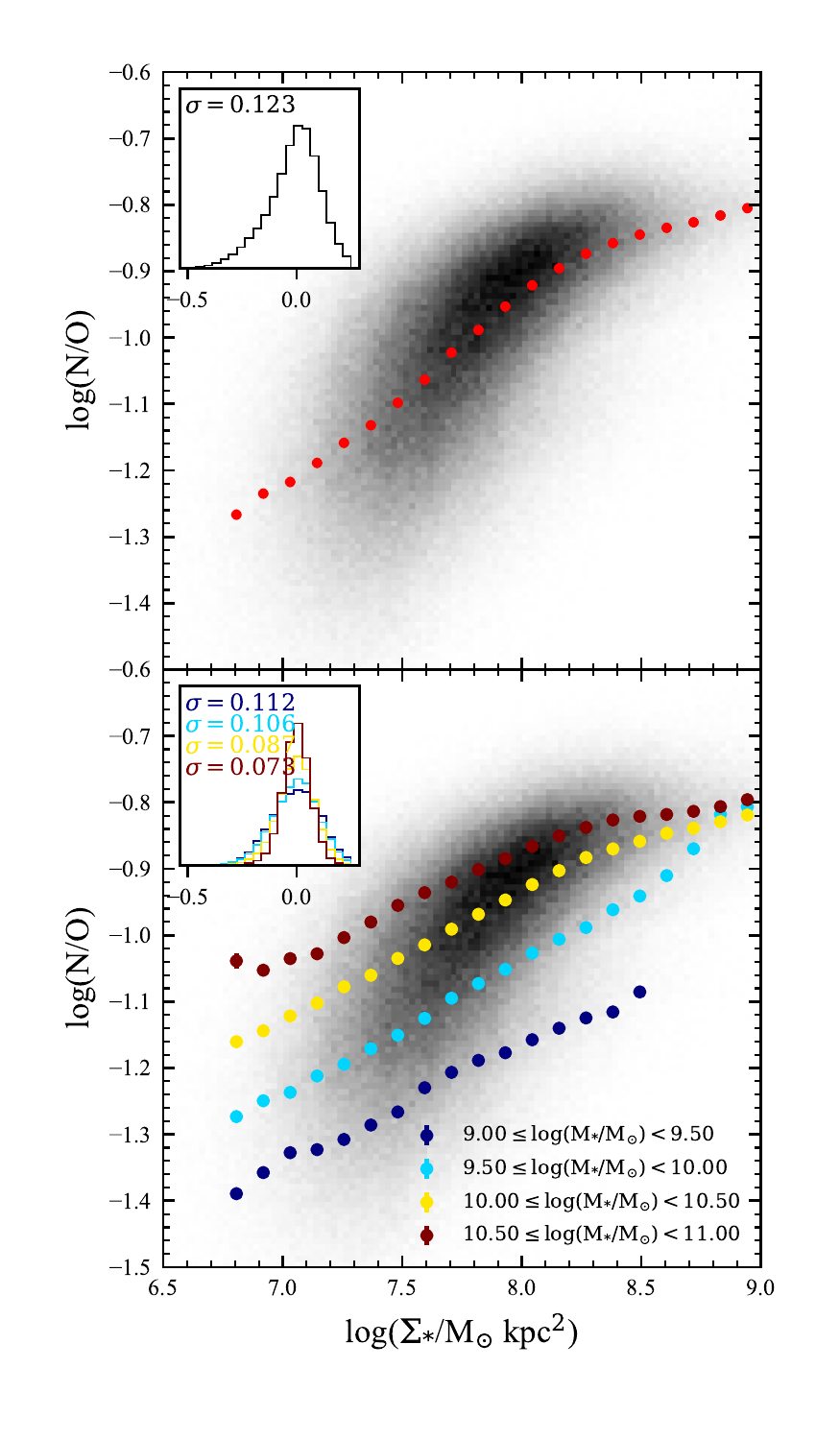}
    \caption{The relationship between N/O and the local stellar mass surface density, $\mathrm{\Sigma_{*}}$. Upper panel: The greyscale shows the two dimensional density of points in this parameter space for all galaxies, while the red points are the median N/O as a function of $\mathrm{\Sigma_{*}}$. In the upper left, we show a normalised histogram of the residuals to a 3rd degree polynomial fit to the medians. Lower panel: The same as the upper panel, but with set of coloured points showing the median N/O in a narrow bin of $\log(\mathrm{\Sigma_{*}}/\mathrm{M_{\odot} \, kpc^{-2}})$ within a range of total stellar mass indicated by the legend. We display the bootstrapped uncertainties on the median values with errorbars, though these are typically smaller than the points. The N/O increases with both local stellar mass density and total stellar mass. In the upper left we show the distribution of residuals around a $3$rd degree polynomial fit to each set of medians. The standard deviation in these residuals is reduced from $\sigma=0.123$ for the full sample to the values shown.}
    \label{NO_sigmastar_split_M}
\end{figure}

Following many previous works on the local oxygen abundances in integral field spectroscopic surveys \citep[e.g.][]{BarreraBallesteros2016}, we now compare N/O to $\mathrm{\Sigma_{*}}$ in our sample. In Figure \ref{NO_sigmastar_split_M}, we show the correlation between N/O and $\mathrm{\Sigma_{*}}$. As is seen with the local oxygen abundance, N/O increases with increasing stellar mass surface density. We calculate the median N/O as a function of $\mathrm{\Sigma_{*}}$ for the full sample and present this in the upper panel of Figure \ref{NO_sigmastar_split_M}. We fit the medians with a third-degree polynomial and then calculate the standard deviation of the residuals to this fit. For our sample, the mean scatter around the N/O-$\mathrm{\Sigma_{*}}$ is $\sigma=0.123$ over the range of stellar densities probed. When we split the sample by total stellar mass in the lower panel of Figure \ref{NO_sigmastar_split_M} and repeat this process, several effects become apparent. The normalisation of the relationship between N/O and $\mathrm{\Sigma_{*}}$ changes with the total stellar mass, such that more massive galaxies have higher N/O at fixed $\mathrm{\Sigma_{*}}$. Secondly, the slope of this relation appears to flatten when looking at narrower bins of stellar mass. The standard deviation of the residuals around the fitted medians is reduced. Thus, the exact shape and scatter in the N/O-$\mathrm{\Sigma_{*}}$ relation will depend on the relative contribution of data points from high and low-mass galaxies.

\subsection{The impact of the star formation rate density}\label{SFRD}
Previous studies of N/O as a function of the integrated stellar mass of galaxies have found no secondary dependence of this abundance ratio on the star formation rate \citep{PerezMontero2013,Kashino2016}. It is not clear that this is also true on local scales, and we will therefore investigate whether it is necessary to take the star formation rate into account to fully understand the local nitrogen abundance.

\begin{figure*}
    \centering
    \includegraphics{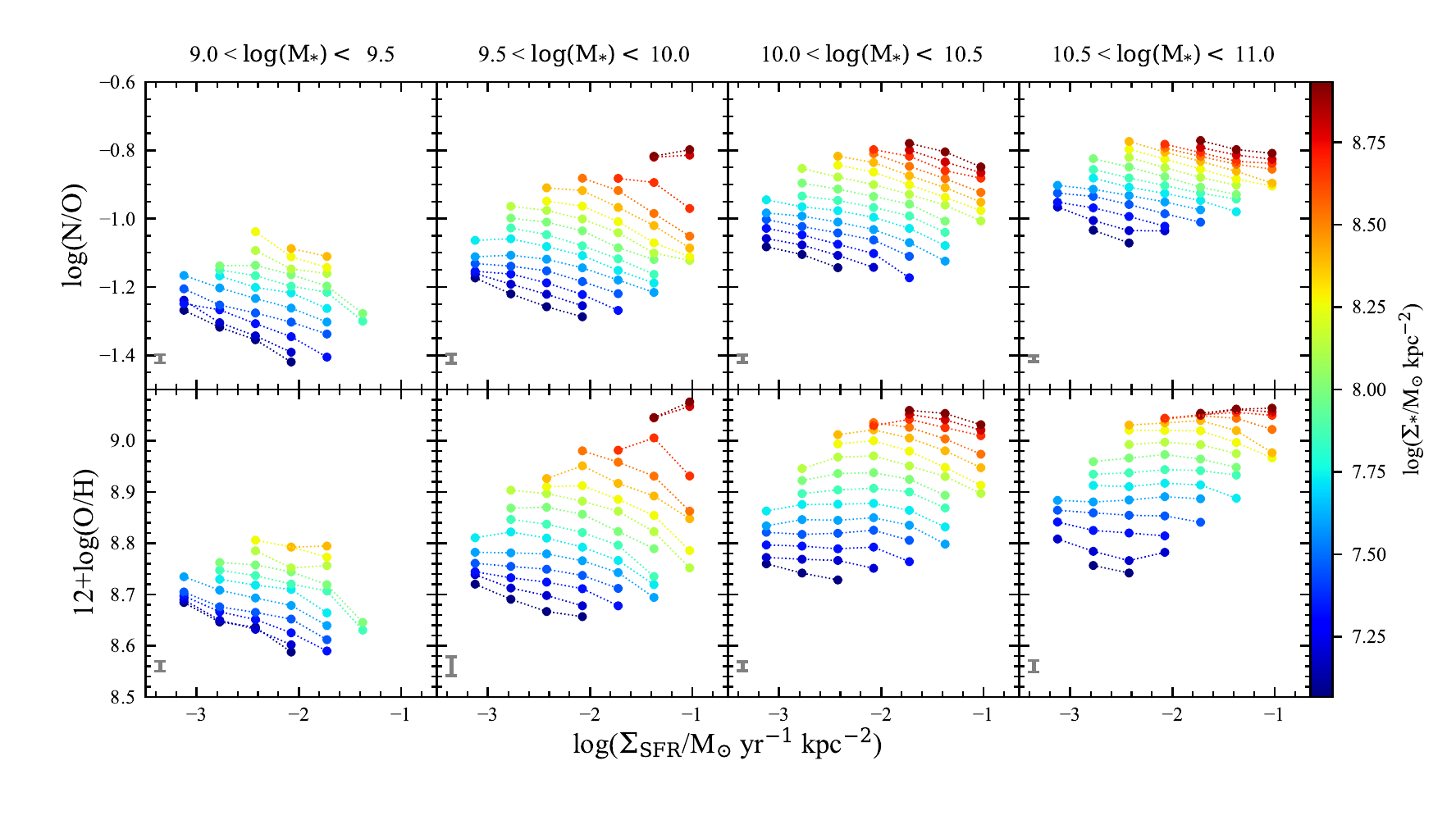}
    \caption{(Upper row) The scaling of the log(N/O) ratio as a function of $\log(\mathrm{\mathrm{\Sigma_{SFR}}})$ and $\log(\mathrm{\Sigma_{*}})$ in bins of total galaxy stellar mass. Each point represents the median log(N/O) in a bin of  $\log(\mathrm{\mathrm{\Sigma_{SFR}}})$ indicated by the abscissa, while the colour of each point represents the stellar mass density as indicated by the colour bar on the right or the range of $\log(\mathrm{\Sigma_{*}}/\mathrm{M_{\odot} \, kpc^{-2}})$ shown in the corresponding colour in the upper left panel. We show only points that are the median of at least $200$ spaxels. The largest bootstrapped uncertainty on any median calculated is displayed at the lower left of each panel. At fixed $\log(M_{*})$ and $\log{\mathrm{\Sigma_{*}}}$, the N/O depends on the star formation rate surface density. (Lower row) The scaling of 12+log(O/H) with $\log(\mathrm{\mathrm{\Sigma_{SFR}}})$ and $\log(\mathrm{\Sigma_{*}})$. This abundance ratio also scales with both $\log(M_{*})$ and $\log{\mathrm{\Sigma_{*}}}$, but its dependence on $\log(\mathrm{\mathrm{\Sigma_{SFR}}})$ is weaker than for log(N/O).
    }
    \label{FMNOR_split_mass}
\end{figure*}

The dependence of the N/O abundance ratio on the star formation rate surface density, stellar mass density and integrated stellar mass is shown in the upper row of Figure \ref{FMNOR_split_mass}. The most important predictors of N/O locally within galaxies are $\log(M_{*})$ and $\log(\mathrm{\Sigma_{*}})$. However, the data also suggest that, in tension with previous studies, the N/O ratio does depend on the star formation rate density.
In order to assess the relative importance of these three parameters on setting the N/O ratio in a given region of the galaxy, we perform a very simple regression, describing the data as
\begin{equation}\label{no_regression}
    \log(\mathrm{N/O})=\beta_{0} + \beta_{1}\log(\mathrm{\Sigma_{*}}) + \beta_{2}\log(\mathrm{\mathrm{\Sigma_{SFR}}}) + \beta_{3}\log(\mathrm{M_{*}}).
\end{equation}

An inspection of the behaviour of N/O as a function of $\log(\mathrm{M_{*}})$ in Figure \ref{NO_sigmastar_split_M}, particularly at high $\mathrm{\Sigma_{*}}$, shows that this model is likely to be insufficient to fully describe the data. Thus the precise results of this regression should be treated with caution. Nevertheless, a Markov-Chain Monte Carlo (MCMC) fit\footnote{MCMC fits of models to the data in this paper made use of the \textsc{lmfit} Python package \citep{Newville2014}, which incorporates the \textsc{emcee} sampler of \cite{ForemanMackey2013}.} for this model to the data finds $\log(\mathrm{N/O})=-5.1 \pm 0.039 +(0.227 \pm 0.0037)\log(\mathrm{\Sigma_{*}}) - (0.094 \pm 0.0034) \log(\mathrm{\mathrm{\Sigma_{SFR}}}) + (0.208 \pm 0.0026) \log(\mathrm{M_{*}})$. The coefficients mean that the dependence of N/O on the local $\mathrm{\mathrm{\Sigma_{SFR}}}$ is roughly half as strong as for $\log(\mathrm{\Sigma_{*}})$ or $\log(\mathrm{M_{*}})$, but importantly it is not zero.

\begin{figure}[ht]
    \centering
    \includegraphics{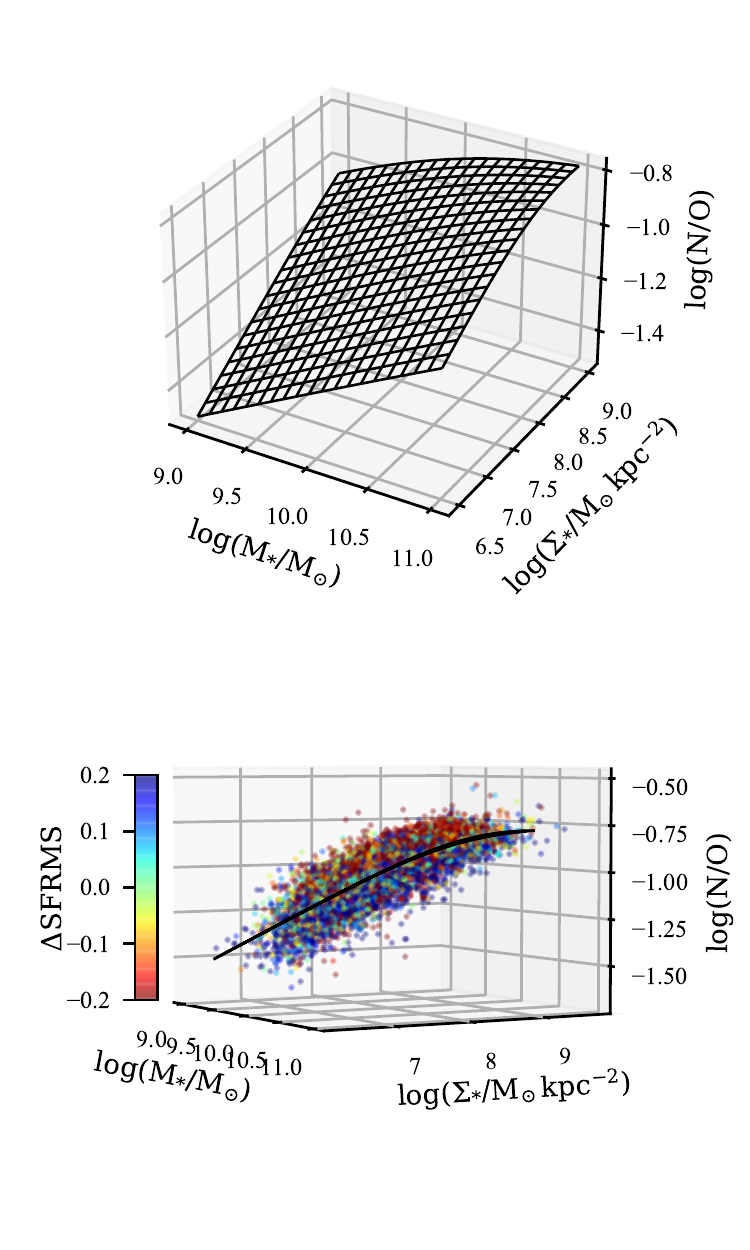}
    \caption{A representation of our fit to the FMNOR given by Equation \ref{no_curti_regression}. In the upper panel we show the surface described by the equation for spaxels that would lie on the resolved star forming main sequence ($\mathrm{\Sigma_{*}}-\mathrm{\Sigma_{SFR}}$ relation). In the lower panel we show a reprojection of the same surface, with the addition of $7927$ randomly selected spaxel measurements representing $1 \%$ of the total sample. These data points are coloured by their residual from the resolved star-formation rate main sequence, clearly showing that spaxels with low $\mathrm{\Sigma_{SFR}}$ for their $\mathrm{\Sigma_{*}}$ tend to have higher log(N/O).}
    \label{3D_FMNOR}
\end{figure}

To more accurately capture the behaviour of N/O at high $\log(\mathrm{M_{*}})$ and $\log(\mathrm{\Sigma_{*}})$, we fit a more flexible functional form to the data. This function incorporates terms that allow for the `turnover' in N/O at high masses and is inspired by Equation (2) of \cite{Curti2020}. Our equation differs by incorporating additional terms for the roughly linear dependence of log(N/O) on $\log(\mathrm{\mathrm{\Sigma_{SFR}}})$ and for the dependence on the local stellar mass surface density,
\begin{equation}\label{no_curti_regression}
    \begin{aligned}
    \log(\mathrm{N/O})=&N_{0} + A\log\left(1+\mathrm{\left(\frac{\mathrm{\Sigma_{*}}}{\Sigma_{0}}\right)^{\beta_{1}} \left(\frac{M_{*}}{M_{0}}\right)}^{\beta_{2}}   \right) \\ + &\beta_{3}\log(\mathrm{\mathrm{\Sigma_{SFR}}}).
    \end{aligned}
\end{equation}
An equation of this form is justified based on the observed flattening in the gradient of the log(N/O)-$\log(\Sigma_{*})$ relation in high stellar mass galaxies. However, this equation is not applicable at low metallicities in the regime where primary nucleosynthesis is the dominant source of nitrogen and log(N/O) reaches a minimum value. This part of parameter space is poorly sampled by our data, so we do not attempt to modify Equation \ref{no_curti_regression} to account for this floor. It is important to note that the interpretation of the $\mathrm{M_{0}}$ and $\Sigma_{0}$ parameters is not as intuitive in this case as in the 1-dimensional case, where they represent a turnover mass. This is due to the fact that the terms involving $\mathrm{\Sigma_{*}}$ and $\mathrm{M_{*}}$ are multiplied together, so the "turnover mass" for one evolves with the other. This behaviour can be seen in the data in the lower panel of Figure \ref{NO_sigmastar_split_M}, where the characteristic $\Sigma_{*}$ where log(N/O) flattens becomes lower at higher $\log(\mathrm{M_{*}}$). We find $N_{0}=-0.85 \pm 0.02$, $A=0.35 \pm 0.05$ $\log(\Sigma_{0})=2.12 \pm 3.13$, $\log(\mathrm{M_{0}})=18.08 \pm 3.59$, $\beta_{1}=-0.76 \pm 0.09$, $\beta_{2}=-0.66 \pm 0.08$ and $\beta_{3}=-0.088 \pm 0.0033$. While the uncertainties on both $\Sigma_{0}$ and $\mathrm{M_{0}}$ are quite high, we note that the posterior distributions from our MCMC show a strong anti-correlation. We will not make any interpretation of these values here, but recognise that a change in one of these parameters in Eq. \ref{no_curti_regression} can be accounted for by a proportionate change in the other variable without changing the estimate of the expected log(N/O). In the fits of both Eq. \ref{no_curti_regression} and Eq. \ref{no_regression}, the derived coefficient for the $\log(\mathrm{\mathrm{\Sigma_{SFR}}})$ term is relatively unchanged. We show a slice of this four dimensional relation in Figure \ref{3D_FMNOR}. The black grid shows the expected log(N/O) for spaxels that sit on the resolved $\mathrm{\Sigma_{*}}$ - $\mathrm{\mathrm{\Sigma_{SFR}}}$ relation, which we find\footnote{To calculate the main sequence we relax the S/N constraints to S/N(H$\alpha$)$>3$ and EW(H$\alpha$)$>3 \, \mathrm{\AA}$ with  H$\alpha/$H$\beta>2.86$. We then perform an ordinary least squares fit to the median $\log(\mathrm{\Sigma_{*}})$ - $\log(\mathrm{\mathrm{\Sigma_{SFR}}})$ with a straight line. This reduces a bias against low $\mathrm{\mathrm{\Sigma_{SFR}}}$ spaxels that would flatten our main sequence estimate.} to have a value of $\log(\mathrm{\mathrm{\Sigma_{SFR}}}/\mathrm{M_{\odot} \, yr^{-1} \, kpc^{-2}}) = 0.81 \log(\mathrm{\Sigma_{*}}/{M_{\odot} \, kpc^{-2}} ) - 8.68$. This is similar, but somewhat shallower than the local main sequence derived by e.g. \cite{Bluck2020}, but we note that a derivation of the precise form of this relation is not the main purpose of this paper. Our conclusions about the SFR-dependence of the local N/O scaling relation are not impacted by our estimate of the local SFR main sequence.

\subsubsection{Comparison to the O/H local scaling relations}
In the lower row of Figure \ref{FMNOR_split_mass} we compare the behaviour of the N/O with that of O/H under variation of the same parameters. In this figure we use the \cite{Kobulnicky2004} nitrogen-free oxygen abundance indicator, based on the R23 and O32 line ratios. With this indicator we see a similar dependence of O/H on the integrated $\log(\mathrm{M_{*}})$ and local $\log(\mathrm{\Sigma_{*}})$, but overall the dependence on $\log(\mathrm{\mathrm{\Sigma_{SFR}}})$ is less pronounced than for log(N/O). Fitted over the entire range of stellar mass with a similar functional form as Equation \ref{no_regression} we find $12+ \log(\mathrm{O/H})=5.95+0.190\log(\mathrm{\Sigma_{*}}) - 0.034 \log(\mathrm{\mathrm{\Sigma_{SFR}}}) + 0.132 \log(\mathrm{M_{*}})$. The dependence of the oxygen abundance on $\log(\mathrm{\Sigma_{SFR}})$ is greater at lower total stellar mass with $12+\log(\mathrm{O/H})\propto -0.045\log(\mathrm{\Sigma_{SFR}})$ in the range $9<\log(\mathrm{M_{*}})<9.5$. For stellar masses in the range $10.5<\log(\mathrm{M_{*}})<11$, we find $12+\log(\mathrm{O/H})\propto -0.013\log(\mathrm{\Sigma_{SFR}})$. The effect of the local star formation rate density on the oxygen abundance is real, but the strength of this effect is a factor $\sim 2 - 4$ weaker than the N/O trend, with the difference changing systematically with the total stellar mass of the galaxies.

The local abundances of nitrogen and oxygen appear to depend on a variety of different local and global properties of galaxies. We will now turn our attention to other galaxy parameters which have an impact on their nitrogen abundances.

\subsection{The effect of galaxy size}
\begin{figure}
    \centering
    \includegraphics{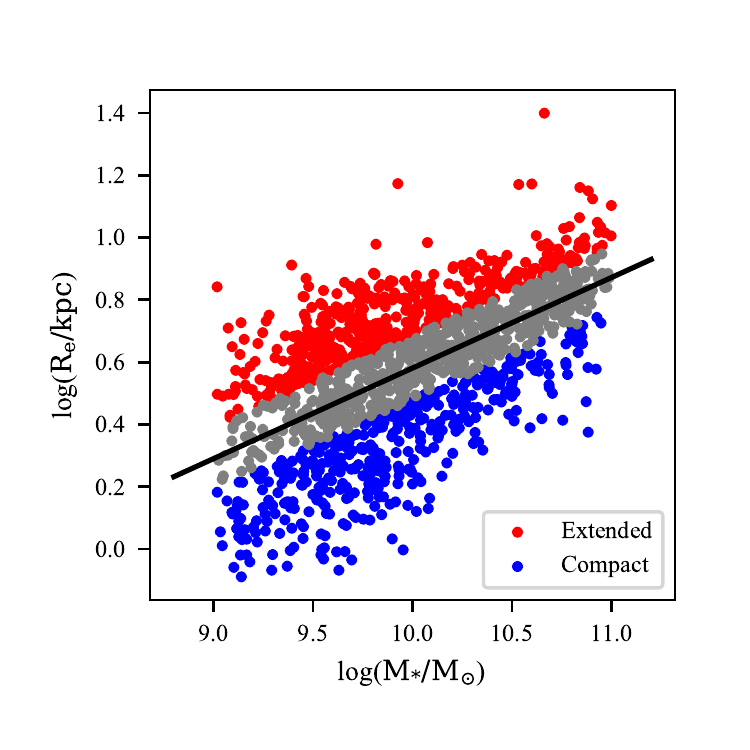}
    \caption{This distribution of the sample on the mass-size diagram. The black line is the result of an unweighted least-squares fit to these data, and has equation $\log(R_{e}/\mathrm{kpc})=0.29\log(\mathrm{M_{*}/M_{\odot}})-2.31$. The line and separates the `extended' subsample (red) from the `compact' subsample (blue), which lie 0.1 dex above and below this line respectively. Grey points are galaxies that lie between the the compact and extended subsamples, and are not included in our analysis of N/O as a function of galaxy size.}
    \label{mass_size_selection}
\end{figure}

Recent work by \cite{Boardman2021} showed that the gas-phase metallicity gradients in galaxies depend both on their total stellar mass as well as their physical size. In this section we show that the details of the local nitrogen scaling relations are also affected by the physical sizes of galaxies, beyond what can be explained by the correlation between stellar mass and size. To illustrate this point we select two subsamples of galaxies based on their position in the mass-size plane. To quantify the size we use the $r$-band elliptical Petrossian half-light radius, which we denote as $R_{e}$. We fit a straight line to the total stellar mass and size, and define galaxies for which the residual is above $0.1$ dex as `extended' and galaxies for which the residual is below $-0.1$ dex as `compact'. The $0.2$ dex gap in $R_{e}$ is intended to eliminate cross-contamination of extended and compact galaxies due to measurement uncertainties and the clustering of the residuals to the fit around zero. While this dramatically reduces the number of spaxels available for the comparison, it ensures that the two subsamples are physically distinct. The definition of the two subsamples is shown in Figure \ref{mass_size_selection}.

The dividing line is described by the equation $\log(R_{e}/\mathrm{kpc})=0.29\log(\mathrm{M_{*}/M_{\odot}})-2.31$, however it is important to stress that the distribution of galaxies in our sample is biased by our selection of star-forming galaxies. The precise definitions of `compact' and `extended' galaxies in this paper are relative and have no profound meaning in the broader context of galaxy evolution.

\begin{figure*}
    \centering
    \includegraphics{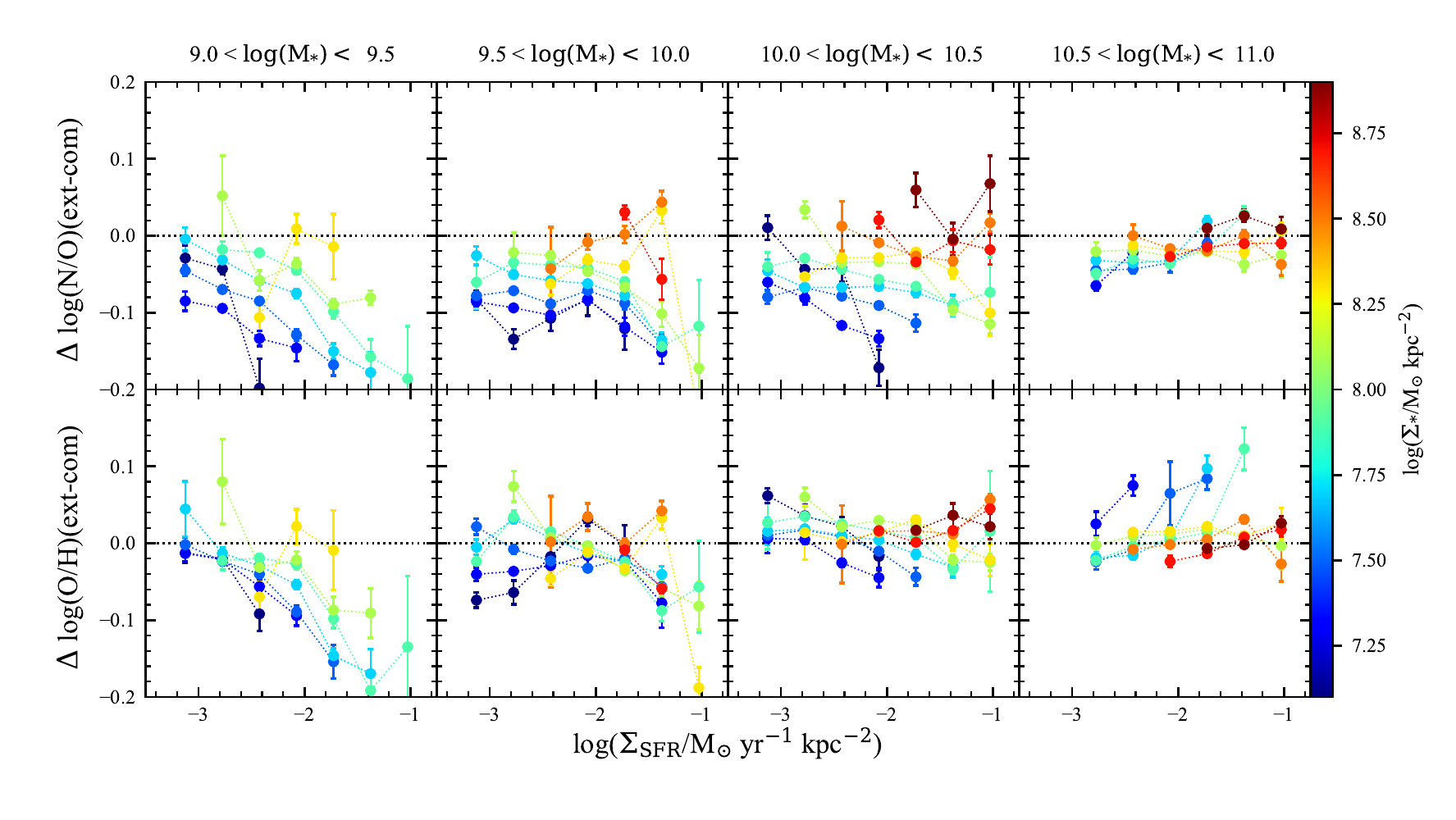}
    \caption{The difference between the local median $\mathrm{\mathrm{\Sigma_{*}}}-{\mathrm{\Sigma_{SFR}}}$-log(N/O) and log(O/H) relations in the extended and compact subsamples. In the top row we show the difference in the median local relations for log(N/O) such that values closer to the top of the plot indicate higher N/O in the extended subsample and lower values indicate higher N/O in the compact subsample. The sizes of the errorbars are calculated by bootstrapping the medians for each subsample and then combining these uncertainties in quadrature. For all stellar masses N/O is enhanced in compact galaxies, especially at low stellar mass surface density. (Bottom row) difference between the local log(O/H) scaling relations for compact and extended galaxies. There is no significant difference in local scaling relations based on galaxy size at a given stellar mass for the \cite{Kobulnicky2004} indicator.}
    \label{delta_no_relation}
\end{figure*}

In Figure \ref{delta_no_relation} we explore the differences in the local log(N/O) and log(O/H) relations between the compact and extended samples defined in Figure \ref{mass_size_selection}. This figure includes data from fewer spaxels than in other parts of this paper due to the reduction in the sample size and the fact that the distributions of $\log(\mathrm{\Sigma_{*}})$ are not identical for the compact and extended galaxies. The two samples show very little systematic difference in their oxygen abundances with the indicator that we have chosen. However, for log(N/O) some differences between the extended and compact samples are visible. The differences are such that the compact galaxies are almost universally higher in N/O both at fixed $\log(\Sigma{*})$ and $\log(\mathrm{\Sigma_{SFR}})$. The difference in N/O is most pronounced for low stellar mass surface density, corresponding to the outskirts of galaxies. Thus in terms of the local scaling relations, the central parts of galaxies are, on average, chemically similar between compact and extended galaxies. It should be noted that since compact galaxies typically have higher central stellar densities, they will tend to have higher central gas-phase metallicities \citep[e.g.][]{Ellison2008}. This size-based offset in the local N/O scaling relations indicates that they are not set up independently of some galaxy-scale process, and that this process is likely to be dependent on either the stellar mass density, the radius, or some related quantity.

\subsection{How local are the N/O scaling relations?}
Given that N/O appears to be related to both local and global effects in galaxies, it is instructive to know which variables play the greatest role in setting this. We investigate this issue using a partial correlation analysis. 

\begin{figure}
    \centering
    \includegraphics{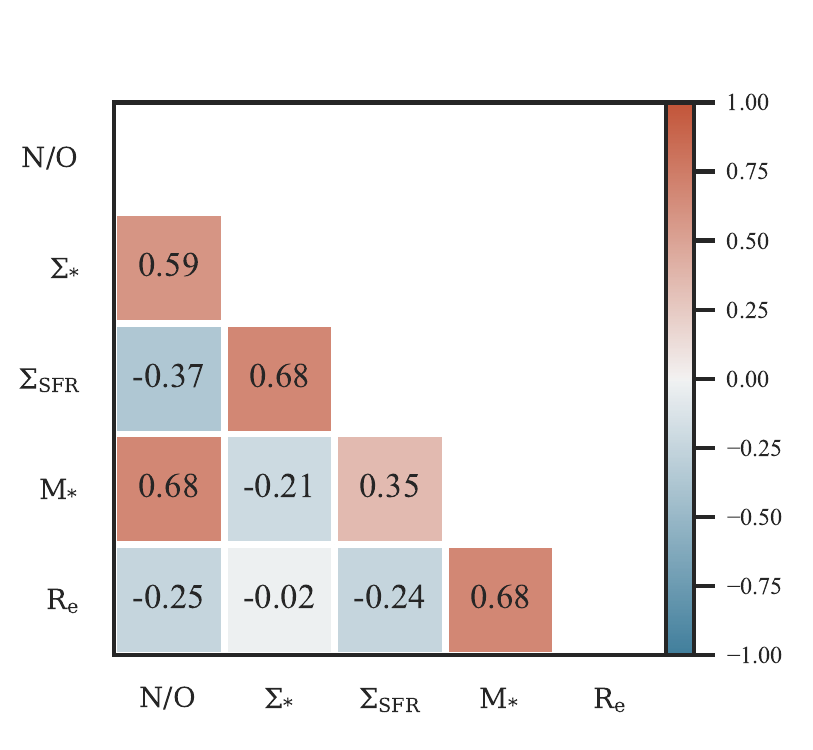}
    \caption{The partial correlation matrix for local and global variable with local N/O. Entries in row $i$ and column $j$ are the Pearson partial correlation coefficient between variables $i$ and $j$, controlling for the effect of all others. The colour of each square shows the strength of the correlation, with red indicating a positive correlation and blue indicating negative. We show the numerical value in each square for clarity. The coefficients are calculated using the log of each quantity. log(N/O) has the strongest partial correlation with $\log(\mathrm{M_{*}})$. }
    \label{partial_correlation_matrix}
\end{figure}

Figure \ref{partial_correlation_matrix} shows a representation of the partial correlation matrix for log(N/O), $\log(\mathrm{\Sigma_{*}})$, $\log(\mathrm{\Sigma_{SFR}})$, $\log(\mathrm{M_{*}})$ and $\log(\mathrm{R_{e}})$. The first column in this matrix shows the partial Pearson correlation coefficient between log(N/O) and the remaining variables. The strongest correlation is between log(N/O) and $\log(\mathrm{M_{*}})$. The sample size is sufficiently large that the $p$-value associated with each partial correlation coefficient is $p<10^{-99}$. These results can therefore be regarded as significant. The squared partial correlation coefficients in the first column of the matrix in Figure \ref{partial_correlation_matrix} tell us the fraction of the total variance in log(N/O) that is explained by each variable, controlling for the effect of others. Thus $\log(M*)$ explains $\sim 46$ per-cent of the variance in N/O when other variables are controlled for, and $R_{e}$ accounts for $\sim 6.3$ per-cent. Variables related to the global properties of galaxies account for approximately half of the variance of log(N/O) in this sample. 

\subsection{The effect of recent star formation history on N/O}
\begin{figure*}
    \centering
    \includegraphics[trim=0cm 0.25cm 0 1.0cm, clip]{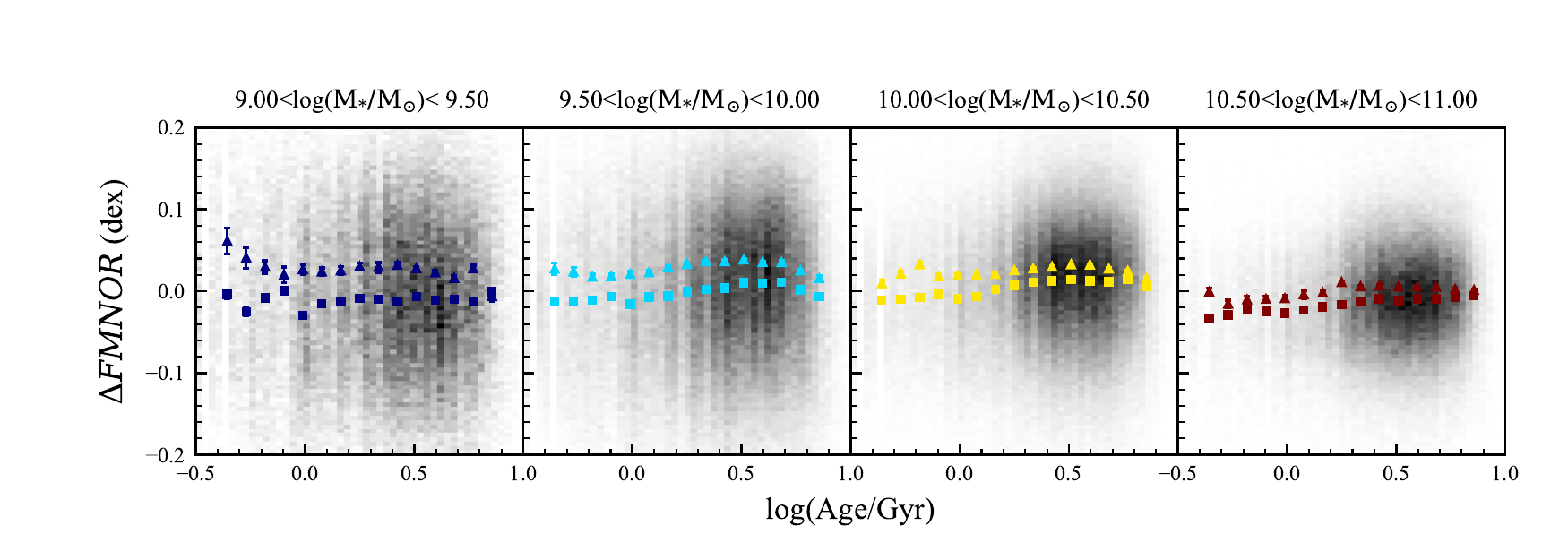}
    \includegraphics[trim=0cm 0.25cm 0 1.0cm, clip]{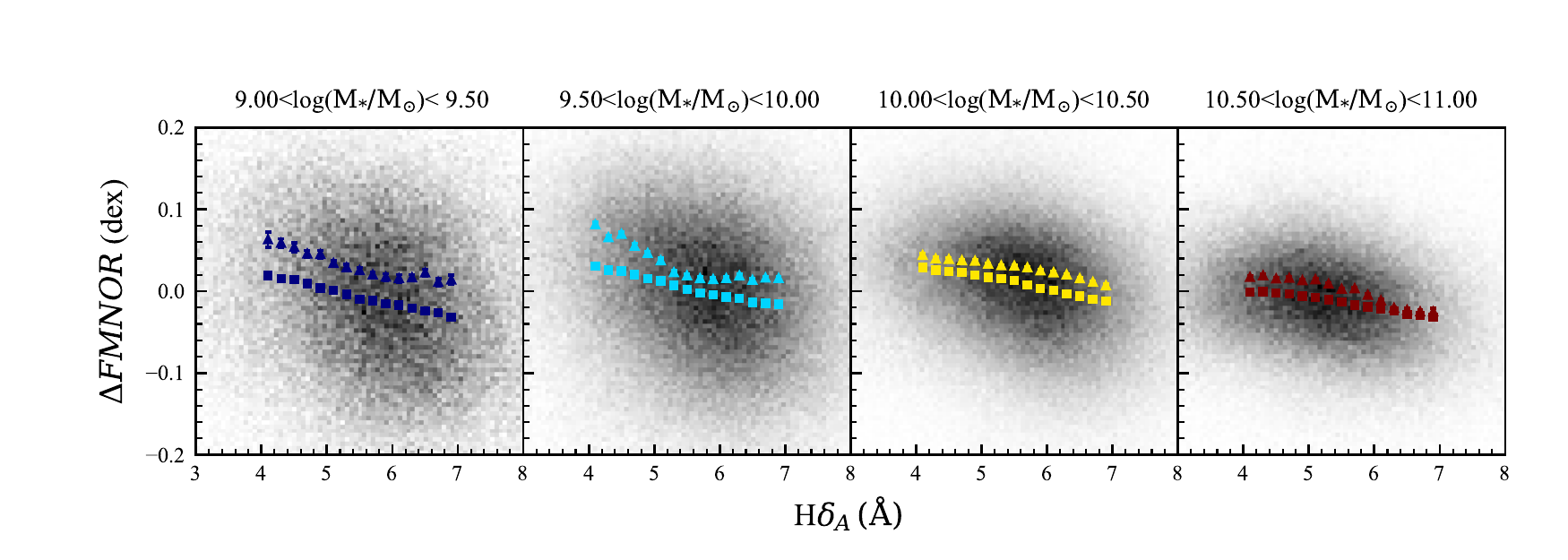}
    \includegraphics[trim=0cm 0.25cm 0 1.0cm, clip]{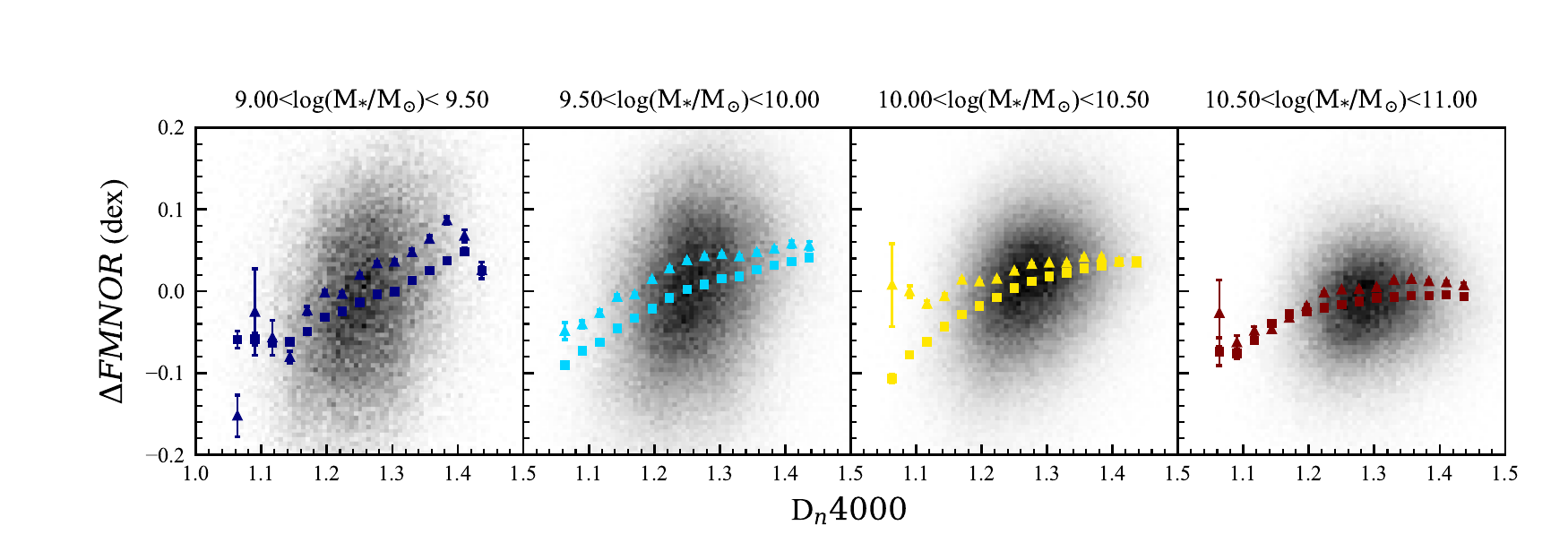}
    \caption{The residuals to the mean log(N/O) - $\log(\mathrm{M_{*}})$ - $\log(\mathrm{\Sigma_{*}})$ - $\log(\mathrm{\Sigma_{SFR}})$ relation as a function of several indicators of stellar population age. The grey shading shows the distribution of residuals for the whole sample in each range of stellar mass, while the points show the median residual for extended galaxies with squares and compact galaxies shown with triangles. (\textit{top row}) light-weighted age as reported by the \textsc{firefly} value added catalogue. The striations are the result of the ages being derived from a set of discrete simple stellar population models; (\textit{second row}) the H$\delta_{A}$ absorption line equivalent width. The evolution of H$\delta_{A}$ is such that age decreases from left to right; (\textit{third row}) D$_{n}4000$. This index increases with increasing stellar population age. Regardless of the stellar population age indicator, the compact galaxies show larger $\Delta$FMNOR than the extended galaxies, indicating an excess of nitrogen at fixed $\log(\Sigma_{*})$, $\log(\Sigma_{SFR})$ and $\log(\mathrm{M_{*}})$. The bootstrapped uncertainties on the median are shown for every point, though are often smaller than the markers.}
    \label{age_residuals}
\end{figure*}

We can investigate the impact of recent star-formation history on N/O in different galaxies by studying how certain spectral features vary with the residuals to the fit presented in Equation \eqref{no_curti_regression}. This approach takes into account the observed trends for log(N/O) with $\mathrm{log(M_{*})}$, $\log(\mathrm{\Sigma_{*}})$ and $\log(\mathrm{\Sigma_{SFR}})$. To do so, we investigate how the residuals to the fit correlate with the strengths of D$_{n}4000$ and H$\delta_{A}$, which are sensitive to variations in the sSFR over timescales of $\sim 100 \, \mathrm{Myr} - 1 \, \mathrm{Gyr}$. The strengths of these features are measured by the MaNGA DAP, using the narrow bandpass definition of \cite{Balogh1999} for D$_{n}4000$ and the prescription of \cite{Worthey1997} for H$\delta_{A}$. As a further check, we extract the light weighted ages for spaxel data from the \textsc{firefly} catalogue \citep{Wilkinson2017,Goddard2017} to compare to the spectral feature measurements. The light-weighted stellar ages are derived from spectra that include a contribution from both young  stellar populations associated with current star formation, and pre-existing older stellar populations. The ages therefore span a large range of values between $1$ and $10$ Gyr. We compare the residuals from the local FMNOR to the various age-sensitive spectral indicators derived from the stellar continuum, split into bins of total $\log(\mathrm{M_{*}})$ in Figure \ref{age_residuals}.

The upper row of Figure \ref{age_residuals} shows no overall correlation of the nitrogen excess above what is expected with the light-weighted age at all stellar ages. There is nevertheless a positive residual in the compact galaxy subsample, the magnitude of which seems to decrease with total stellar mass. The residual trends with the stellar features H$\delta_{A}$ and D$_{n}4000$ also show  an almost constant offset between the extended and compact galaxies. Interestingly, the median residuals for both subsamples appear to increase with increasing stellar age (recall that lower D$_{n}4000$ corresponds to younger stellar ages, while lower H$\delta_{A}$ corresponds to older stellar ages). The effect is largest for galaxies with the lowest stellar masses ($9<\log(\mathrm{M_{*}/M_{\odot}})<9.5$), but is also present for galaxies above $\log(\mathrm{M_{*}/M_{\odot}})=10.5$. At the low $\log(\mathrm{M_{*}})$ end, $\Delta$FMNOR drops by $0.05$ dex across the range H$\delta_{A}=4-7 \, \mathrm{\AA}$. At the highest total stellar masses studied $\Delta$FMNOR drops by $0.03$ dex over the same range in H$\delta_{A}$. Similarly, $\Delta$FMNOR varies by up to $0.13$ with D$_{n}4000$ in the lowest total stellar mass interval over the range $1.1<\mathrm{D}_{n}4000 <1.45$, and by $0.075$ dex in the highest stellar mass interval. 

Thus, even if the local instantaneous $\mathrm{\Sigma_{SFR}}$ and $\mathrm{\Sigma_{*}}$ are taken into account, N/O appears to increase with the local stellar age as traced by D$_{n}4000$ and H$\delta_{A}$. The discrepancy between the modelled light-weighted age and the measured spectral features may be due to a secondary dependence of D$_{n}4000$ and H$\delta_{A}$ on the metallicity of the stellar population. Regardless of the differences in the median behaviour of the N/O residuals with these age indicators, it is clear that the local properties of the stellar populations cannot explain the difference in gas-phase chemical abundances between extended and compact galaxies.

\subsection{Is the relationship between N/O and O/H universal?}
\begin{figure*}
    \centering
    \includegraphics{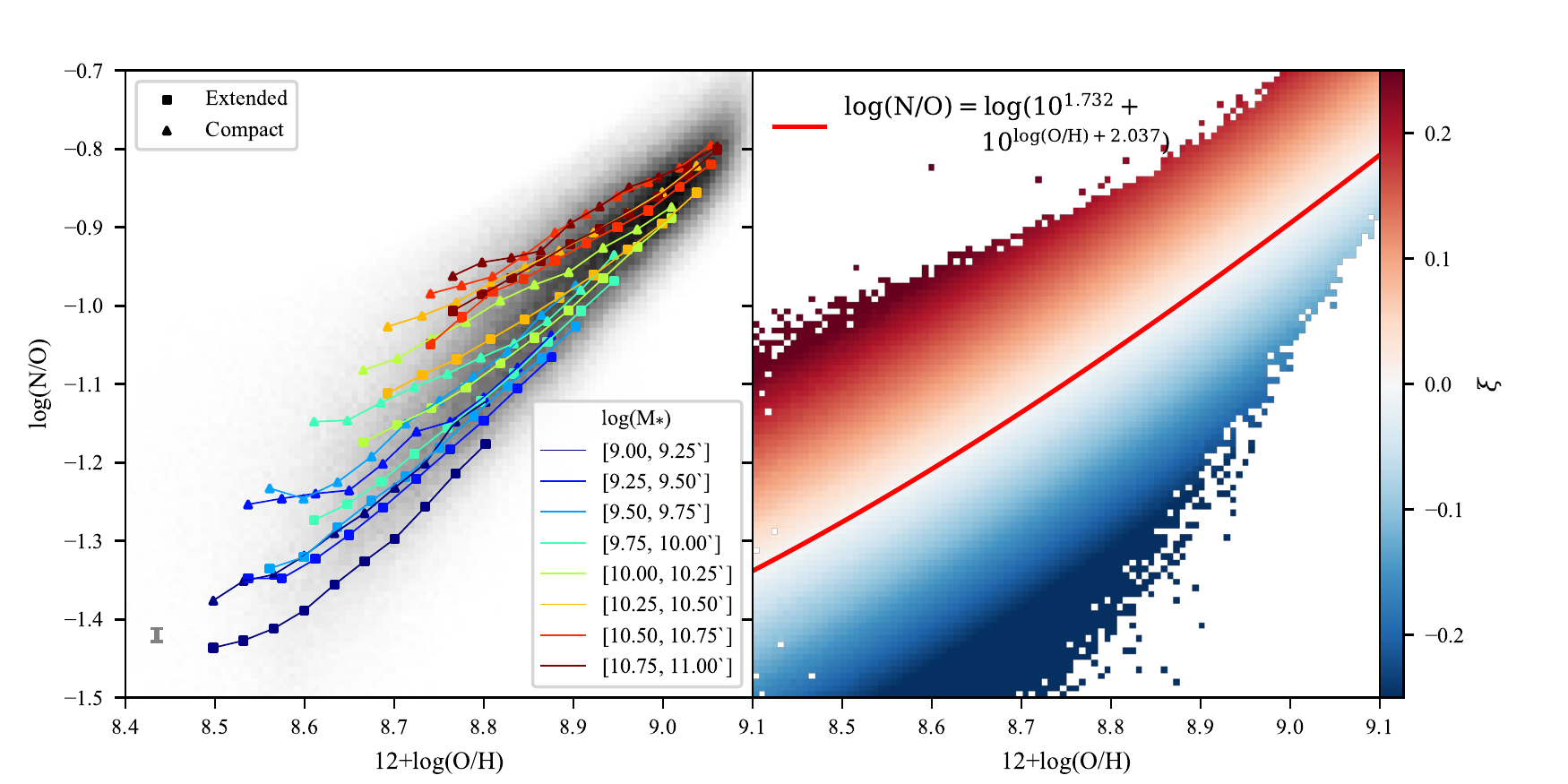}
    \caption{(Left) The median log(N/O) as a function of 12+log(O/H) in galaxies of different total stellar mass and size. Lines of the same colour have the same integrated stellar mass. Lines marked with a square come from the extended subsample, while lines marked with a triangle come from the compact subsample. At a given stellar mass, the median N/O is always higher in the compact subsample than in the extended subsample at a fixed 12+log(O/H). The grey errorbar at the lower left shows the largest bootstrapped uncertainty on any median plotted.  (Right) The definition of the nitrogen excess factor, $\xi$. The red line shows the fitted N/O-O/H relation of the form described by \cite{Nicholls2017}, which is $\log(\mathrm{N/O}) = \log(10^{-1.732} + 10^{\log(\mathrm{O/H})}+2.037)$. The colourmap shows $\xi$, the residuals of our data from this line.}
    \label{NO_OH_comp_ext}
\end{figure*}

The differences in the local scaling relations between galaxies of different mass and size demand further exploration. The differing behaviour of N/O and O/H across the mass-size plane suggests that the relationship between N/O and O/H may not be universal in our sample. This is explored in Figure \ref{NO_OH_comp_ext}, where we show the N/O-O/H relation as a function of integrated stellar mass, showing the median N/O as a function of O/H for the extended and compact subsamples. Since our sample includes only galaxies with $\log(\mathrm{M_{*}})>9.0$, our data cover the range of metallicity in which secondary nitrogen dominates, and we are unable to explore any systematics in the primary N/O ratio. A similar decomposition has been explored previously in \cite{Belfiore2017} and \cite{Schaefer2020}, however the effect of selecting galaxies of different sizes at fixed stellar mass has, to our knowledge, not been studied before.

As has been noted in earlier works \citep{Belfiore2017,Schaefer2020}, the slope and normalisation of the N/O and O/H relation depends on the total stellar mass of the galaxies observed, with higher mass systems having higher N/O at fixed O/H, and a shallower relationship between these two ratios. At all stellar masses, however, the relative sizes of galaxies influence the N/O-O/H relation such that more compact galaxies are more nitrogen-rich. The differences become more prominent at lower $\log(\mathrm{M_{*}})$ and at lower values of 12+log(O/H). For galaxies with $\log(\mathrm{M_{*}})>10.5$, N/O is $\sim 0.05$ dex higher in compact galaxies, while below this, the difference is as high as $0.1$ dex. The divergent behaviour leads to more scatter in the N/O-O/H relation at lower metallicities. The systematic discrepancies are most visible at $12+\log(\mathrm{O/H})\sim 8.8$, where there is overlap in the metallicity distributions from the highest and lowest mass galaxies in our sample. At this oxygen abundance, there is $~0.25$ dex systematic offset in the N/O ratio between the most massive compact galaxies and the least massive extended galaxies. 

We can quantify the difference in the N/O-O/H relation between galaxies across the mass-size plane by fitting linear functions to the median relations shown in Figure \ref{NO_OH_comp_ext}. At low metallicity, below $12+\log(\mathrm{O/H})\sim 8.6$ with the KK04 indicator, the relation reaches a floor. Our data do not sample such low metallicities across the entire stellar mass range, so we restrict this analysis to the regime dominated by secondary nitrogen. We fit the medians for $12+\log(\mathrm{O/H}) > 8.6$ with an equation of the form
\begin{equation}\label{no_oh_linear}
    \log(\mathrm{N/O})=ax+b,
\end{equation}
where $x=12+\log(\mathrm{O/H})-8.7$, and $b$ is the value of log(N/O) where $12+\log(\mathrm{O/H})=8.7$. The fitted coefficient values in these equations are presented in Table \ref{no_oh_coeff}. In agreement with what can be seen qualitatively in Figure \ref{NO_OH_comp_ext}, the gradients for the N/O-O/H relation quantitatively flatten toward higher stellar masses and in more compact galaxies. The values of log(N/O) at $12+\log(\mathrm{O/H})=8.7$ similarly increase in more massive and more compact galaxies.

\begin{table*}[]
    \centering
    \begin{tabular}{c|c|c||c|c}
        $\log(\mathrm{M_{*}})$& $a_{ext}$ & $b_{ext}$ & $a_{com}$ & $b_{com}$ \\ 
        \hline
        \hline
        9.00 - 9.25&$1.075 \pm 0.041$ &$-1.290 \pm 0.002 $&$1.032 \pm 0.056 $&$-1.228 \pm 0.003$ \\
        9.25 - 9.50&$0.984 \pm 0.015$ &$-1.242 \pm 0.001 $&$0.781 \pm 0.045 $&$-1.186 \pm 0.004$ \\
        9.50 - 9.75&$0.973 \pm 0.012$ &$-1.226 \pm 0.001 $&$0.926 \pm 0.023 $&$-1.167 \pm 0.003$ \\
        9.75 - 10.00&$0.933 \pm 0.023$ &$-1.206 \pm 0.003 $&$0.626 \pm 0.044 $&$-1.116 \pm 0.006$ \\
        10.00 - 10.25&$0.849 \pm 0.034$ &$-1.164 \pm 0.006 $&$0.610 \pm 0.010 $&$-1.064 \pm 0.002$ \\
        10.25 - 10.50&$0.732 \pm 0.024$ &$-1.117 \pm 0.005 $&$0.591 \pm 0.019 $&$-1.033 \pm 0.004$ \\
        10.50 - 10.75&$0.690 \pm 0.018$ &$-1.068 \pm 0.004 $&$0.584 \pm 0.019 $&$-1.006 \pm 0.004$ \\
        10.75 - 11.00&$0.670 \pm 0.015$ &$-1.051 \pm 0.003 $&$0.594 \pm 0.031 $&$-1.012 \pm 0.007$ \\
    \end{tabular}
    \caption{The fitted coefficients of log(N/O)$=ax+b$, where $x=12+\log(\mathrm{O/H})-8.7$, for the extended and compact subsamples of galaxies. These were only fitted for datapoints with $12+\log(\mathrm{O/H})\geq 8.6$, in the regime where secondary nitrogen production dominates. The slope of the N/O-O/H relation decreases with increasing total stellar mass, while the value of log(N/O) where $12+\log(\mathrm{O/H})=8.7$ increases. In a given stellar mass range the N/O-O/H relation is flatter for compact galaxies and the constant term is higher.}
    \label{no_oh_coeff}
\end{table*}

\begin{figure*}
    \centering
    \includegraphics{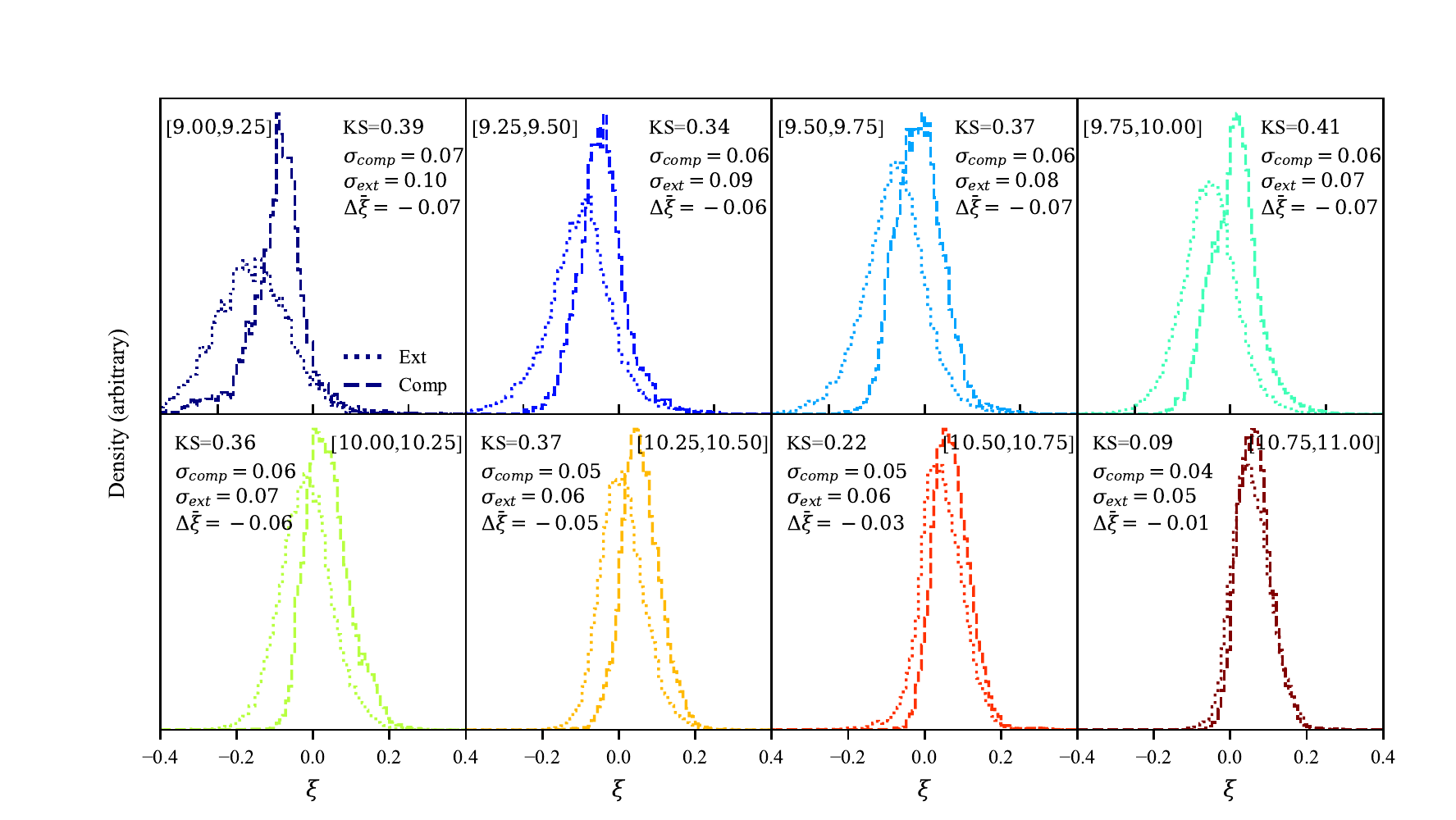}
    \caption{The distributions of the nitrogen excess factor, $\xi$, for compact galaxies (dashed lines) and extended galaxies (dotted lines). Each panel shows a different range of $\log(\mathrm{M_{*}})$ shown in square brackets. We report the Kolmogorov-Smirnov statistic, which all have p-value lower than $10^{-42}$, the standard deviations of $\xi$ in compact ($\sigma_{comp}$) and extended ($\sigma_{ext}$) galaxies, and the difference between the mean of each distribution $\Delta\bar{\xi}$. In each mass range, the difference between $\xi$ in each subsample is highly significant.}
    \label{Xi_hists}
\end{figure*}

Systematic differences the N/O abundance scaling relations with parameters such as galaxy size will be important for characterising the N/O-O/H relation in surveys of H{\sc ii} regions in local galaxies. While the mean relation between N/O and O/H may vary between galaxies of different sizes, it is important to know how large these differences are relative to the scatter and whether the distributions of chemical abundances are truly different. We visualise this in Figure \ref{Xi_hists}, where we show the distributions of the parameter $\xi$ for compact and extended galaxies in each interval of total stellar mass. As in \cite{Schaefer2020}, we define $\xi$ as the residual of each N/O estimate from the mean N/O-O/H relation as determined by a fit of a function of the form suggested by \cite{Nicholls2017} to the data. Namely, we find the relation $\log(\mathrm{N/O})=\log(10^{-1.732} + 10^{\log(\mathrm{O/H}) + 2.037})$, where we have taken the lower floor of log(N/O)$=-1.732$ as determined by \cite{Nicholls2017}, since we lack sufficient low-metallicity data to constrain this parameter. The definition of $\xi$ is shown graphically in the right hand panel of Figure \ref{NO_OH_comp_ext}.
To evaluate the difference in the distributions of $\xi$ between extended and compact galaxies we calculate the two-sample Kolmogorov-Smirnov \citep[KS][]{Kolmogorov1933} statistic in each mass bin, as well as the difference in the means, and the standard deviation of each distribution. These are shown in each panel in Figure \ref{Xi_hists}. The KS statistic is generally larger at lower stellar masses, but reaches a maximum of $0.41$ in the range $9.75<\log(\mathrm{M_{\odot}})<10.0$. The KS statistic is highly significant in all mass ranges, with computed p-values smaller than $10^{-42}$.

We also report the differences in the means of each distribution, $\Delta\bar{\xi}$, and their standard deviations. For $\log(\mathrm{M_{*}})<10.5$, $\Delta\bar{\xi}$ is similar in magnitude to the standard deviation of one of the distributions of $\xi$, meaning that the mean N/O for compact galaxies in one mass-range is typically $\sim 1 \sigma$ higher than in extended galaxies. With each distribution containing many thousands of points, the nitrogen abundances in extended galaxies is therefore significantly different than in compact galaxies.

\section{Discussion}\label{Discussion}

\subsection{Does a Fundamental Mass N/O Relation exist on local scales?}
Given the arguments that the global mass-metallicity relation arises from a more fundamental local relationship between stellar mass density and oxygen abundance \citep{RosalesOrtega2012,BarreraBallesteros2016,Sanchez2020ARAA}, it is of interest to explore whether something similar holds for N/O. Where previous authors \citep[e.g.][]{PerezMontero2013, Andrews2013} found a tight correlation between $\log(\mathrm{M_{*}})$ and N/O, with no secondary dependence on SFR, we find that log(N/O) decrease by $0.088-0.094$ dex per dex increase in $\Sigma_{SFR}$. Moreover, since our data show that the N/O measured in galaxies is moderated by a combination of both local stellar mass density and global stellar mass, we are unable to explain the lack of a global SFR trend reported by \cite{PerezMontero2013} by local properties alone.

The offsets in the N/O scaling in Figure \ref{delta_no_relation} show that this size effect cannot simply be attributed to differing distributions of $\mathrm{\Sigma_{*}}$ or $\mathrm{\Sigma_{SFR}}$ between the compact and extended samples. 

There are a number of possible explanations for this effect \citep[see e.g.][]{Belfiore2015}. 
\begin{enumerate}[(i)]
    \item{Bursts of star formation resulting in changes in the star formation efficiency without gas accretion. In this scenario, which may result from galaxy-galaxy interactions, a burst of star formation will initially cause a decrease in N/O and in increase in O/H, but an increase in N/O later as the evolution of low and intermediate mass stars progresses \citep{Garnett1990}}
    \item{The accretion of low-metallicity gas, coupled with the delayed production of nitrogen will decrease O/H without changing N/O, moving galaxies above the average N/O-O/H relation \citep{Koppen2005}. The accretion of gas is likely to result in a corresponding increase in $\mathrm{\Sigma_{SFR}}$ due to the Kennicutt-Schmidt law \citep{Schmidt1959,Kennicutt1998}. } \label{accretion}
    \item{If extended and compact galaxies differ systematically in their recent star formation histories, then the relative abundances of N and O will differ between these two subpopulations due to the delayed production of nitrogen.}\label{sfh}
    \item{Galactic fountain flows, whereby feedback moves enriched gas from the centres of galaxies to their outskirts \citep{Shapiro1976}. This will increase N/O in the outer parts of galaxies, with the precise level of enhancement depending on the abundances in the fountain flow.}\label{fountains}
    \item{Differential outflows, where the relative proportions of N and O expelled by stellar feedback is not fixed but may instead depend on the gravitational potential and the details of how galactic winds are launched \citep{Vincenzo2016}. }\label{diff_out}
    \item{The growth of a bulge has been linked to the quenching of star formation in galaxies \citep{Fang2013}. The bulge that results from this phase of `compaction' would have the effect of stabilising the galaxy's gaseous disk and reducing the star formation efficiency \citep{Martig2009}. Models show that this reduced efficiency can result in increasing N/O at fixed O/H \citep[e.g.][]{Molla2006}.  }\label{morph_quenching}
\end{enumerate}
These processes are all likely to occur in the Universe to some degree.
\cite{Schaefer2020} showed that within galaxies, radial variation in star formation efficiency is correlated with deviations of N/O from what is expected given the measured O/H. However, the radial variations in the inferred star formation efficiency could not completely explain the range of N/O at fixed O/H in that data set.
Indeed, there is observational evidence of process (\ref{accretion}) occurring in MaNGA \citep{Luo2021} in anomalously low metallicity regions, though \cite{Hwang2019} report that such regions account for only $\sim 10$ percent of star-forming spaxels in MaNGA. Given that we measure a decline in N/O at high star formation rate in the medians in Figure \ref{FMNOR_split_mass}, we conclude that this process is probably not the dominant factor in producing our observed trends, nor can it account for the differences in the relationship between N/O and O/H across the mass-size plane.

Our results do provide evidence that the recent star formation history has some effect on the local N/O ratios. The correlation of log(N/O) with $\Sigma_{SFR}$ shown in Figure \ref{FMNOR_split_mass}, as well as the correlations of $\Delta$FMNOR with D$_{n}4000$ and H$\delta_{A}$ in Figure \ref{age_residuals} suggest that the delayed production of nitrogen has a measurable impact on the abundance ratios in galaxies. However, local star formation histories are unable to reconcile the difference between the log(N/O) in compact and extended galaxies. It may be true that reductions in the star formation efficiency over long timescales are responsible for the offset between compact and extended galaxies, such as in point \ref{morph_quenching} above. If this is so, then the differences in log(N/O) shown in Figure \ref{age_residuals} must have been established well over a Gyr ago, since they persist even in spaxels with the oldest light-weighted ages. 

We have argued against all the items in the above list except for \ref{fountains} and \ref{diff_out}, that is the redistribution of nitrogen through fountain flows, and the differential loading of nitrogen and oxygen in outflows. In practice these two processes are probably not completely independent, and the chemical abundances in fountain flows may be related to the abundances in outflows, though this will depend on the mechanisms by which these flows are launched. In compact galaxies the star formation rate densities are typically higher, given the relationship between $\mathrm{\Sigma_{*}}$ and $\mathrm{\Sigma_{SFR}}$, and have higher central log(N/O). Should a fountain flow be driven, more nitrogen-rich gas will be redistributed to the outer parts of the galaxy. This will have the effect of driving up the local log(N/O) scaling relations relative to more extended galaxies. It should be noted that a number of theoretical works cast doubt on the ability of galactic fountains to impact the chemical abundances over large scales. Models of fountains in the Milky Way have shown that the typical distance over which material is dispersed by galactic fountains is of order $\sim 0.5$ kpc \citep{Bregman1980,Fraternali2008}. Consequently, the metallicity gradients in the Milky Way are unlikely to be strongly affected by fountain flows \citep{Melioli2008,Melioli2009,Spitoni2009}. These models deal with Milky Way mass galaxies, which would sit in our upper stellar mass bins \citep{Licquia2015} where we see the smallest difference between the compact and extended subsamples. Moreover, they do not explicitly report on the evolution of nitrogen. For this reason it is difficult to completely rule out the possibility of galactic fountains playing a role in the N/O trends seen in our data. A thorough treatment of this topic is clearly necessary. This will require a careful comparison of the data to simulations. This is beyond the scope of this work and we defer this endeavour to a future paper.

\subsection{Implications for observational abundance studies}
Regardless of the physical origin of the systematic differences in the N/O-O/H relation in galaxies of different kinds, these differences have implications for the calibration of strong line abundance indicators. Previous works \citep{PerezMontero2009,Schaefer2020} have pointed out that since many strong-line O/H abundance indicators assume a fixed N/O-O/H relation, some strong line indicators will be systematically biased under some circumstances. Since the strength of the [NII]$\lambda6584$ line is proportional to the absolute nitrogen abundance, an error in the N/O-O/H relation used to calibrate these metallicity diagnostics will lead to systematic uncertainties in the overall oxygen abundance scale. Implicit in Figure \ref{NO_OH_comp_ext} is the fact that the precise form of the N/O-O/H relation observed will be subject to selection effects. The relation measured in galaxy centres will be different from in their peripheries, and the relation derived from Local Group H{\sc ii} regions may not be applicable in other parts of the Universe. A similar effect has been seen using direct method elemental abundances, where an offset in N/O at fixed O/H has observed between local star-forming galaxies and nearby high-z analogues \citep[e.g.][]{Bian2020,PerezMontero2021}. Between these findings and the results of our current work, we must conclude that the N/O-O/H relation is far from universal, and that the assumptions about the relative abundances of N and O underpinning many strong line indicators are not strictly true. More work is needed to determine the possible impact of H{\sc ii} region sample selection on empirically-derived abundance indicators.

\section{Conclusions}\label{Conclusions}
The idea that the chemical abundances in galaxies are set on local scales is an attractive one. It has the power to explain not only the global mass-metallicity relation, but also the metallicity gradients within galaxies. In this paper we have tested this idea by exploring the relationship between log(N/O) and the local and global properties of galaxies. For the first time we have presented the relationship between log(N/O) and $\mathrm{\Sigma_{*}}$, finding a strong positive correlation. However, the data strongly suggest that local variables alone are insufficient to explain log(N/O).
In summary we find that:
\begin{itemize}
    \item{When controlling for other variables the total stellar mass of galaxies is most strongly correlated with the \emph{local} log(N/O), with $r=0.68$. The correlation of log(N/O) with local $\log(\mathrm{\Sigma_{*}})$ controlling for total stellar mass is slightly lower with $r=0.59$. Thus, log(N/O) is not only related to the integrated star-formation history locally within galaxies, but is also associated with the total stellar content of the galaxy as a whole. Observationally, this suggests that studies of the local chemical abundances in galaxies cannot be made in a way that is agnostic to the total stellar mass distribution in the sample.  }
    \item{log(N/O) is significantly correlated with the local star formation rate surface density, $\log(\mathrm{\Sigma_{SFR}})$ with Pearson's $r=-0.37$ controlling for $\log(\mathrm{M_{*}})$, $\log(\mathrm{\Sigma_{*}})$ and $\log(\mathrm{r_{50}})$. Two separate regression models presented in Equations \ref{no_regression} and \ref{no_curti_regression} find that log(N/O) decreases by $0.088-0.094$ for each dex increase in $\mathrm{\Sigma_{SFR}}$. This is in agreement with the simulation predictions for whole galaxies of \cite{Matthee2018}, who attributed this trend to the delayed production of nitrogen, but is at odds with the observations of \cite{PerezMontero2013}.}
    \item{We observe that at fixed $\log(M_{*})$ and $\log(\Sigma_{*})$, log(O/H) decreases by $0.034$ per dex increase in $\Sigma_{SFR}$ when averaged over the whole sample. In the range $9<\log(M_{*})<9.5$, log(O/H) decreases by $0.045$ per dex increase in $\Sigma_{SFR}$, but for galaxies with $10.5<\log(M_{*})<11$, this dependency is reduced to a $0.013$ reduction in log(O/H) per dex increase in $\Sigma_{*}$. At the same time, we observe that $\log(\mathrm{N/O})\propto (0.088-0.091)\log(\Sigma_{SFR})$, depending on how we account for other variables in our regression model. Thus the effect of $\log(\mathrm{\Sigma_{SFR}})$ on log(N/O) is approximately $2$-$3$ times greater than its effect on log(O/H). This may mean that detections of a SFR-dependence of the local $\mathrm{\Sigma_{*}}$-O/H relation using N-based strong line calibrations will overestimate the size of the effect.}
    \item{In addition to the dependence of log(N/O) on $\Sigma_{SFR}$, we find that the residuals to the FMNOR fit to Equation \ref{no_curti_regression}, which includes a term for the instantaneous $\Sigma_{SFR}$, correlate with D$_{n}4000$ and H$\delta_{A}$. Controlling for $\log(\mathrm{M_{*}})$, $\log(\Sigma_{*})$ and $\log(\Sigma_{SFR})$, log(N/O) will increase by between $0.075 - 0.13$ dex between $1.1<\mathrm{D}_{n}4000<1.45$, and decrease by $0.03-0.05$ dex between $4<\mathrm{H}\delta_{A}<7 \, \mathrm{\AA}$. These trends vary systematically with total stellar mass such that the strongest trends are found at low stellar mass. Since D$_{n}4000$ and H$\delta_{A}$ are sensitive to star formation rate timescales that are longer than for H$\alpha$ emission ($\sim 100 \,  \mathrm{Myr}$ for D$_{n}4000$ and H$\delta_{A}$ compared to $\sim 10 \, \mathrm{Myr}$ for H$\alpha$ emission) , this indicates that the delayed production of nitrogen has a measurable effect on the chemistry of galaxies.}
    \item{We confirm the total stellar mass dependence of the N/O-O/H relation \citep{Belfiore2017,Schaefer2020}, and report an additional dependence on the galaxy size at fixed stellar mass. More compact galaxies are observed to be consistently more nitrogen-rich. The size difference is most pronounced in lower stellar mass galaxies. Below $\log(M_{*})=9.5$ we observe log(N/O) to be $0.075$ higher at all metallicities present in galaxies in this mass range. Meanwhile, for galaxies above $\log(M_{*})=10.5$, the difference in log(N/O) at fixed log(O/H) is reduced to $0.03$. The variation in log(N/O) cannot be explained by invoking a universal N/O-$\mathrm{\Sigma_{*}}$-$\mathrm{\Sigma_{SFR}}$ relation and argues against a universally applicable N/O-O/H relation. This difference is largest at low stellar mass and furthermore cannot be attributed to differences in the \emph{local} star formation history. The chemical abundance patterns across galaxies are influenced by their total mass and morphology.}
\end{itemize}

It is important to note that the exact magnitude of the trends we have reported, as well as the differences between subpopulations of galaxies, is sensitive to the choice of strong-line indicator that we have employed. This has been highlighted in Appendix \ref{Different_indicators}. While the sizes of differences between extended and compact galaxies varies between indicators, the scatter in the scaling relations proportionally. Thus, the significance of the differences remains, regardless of the indicator employed.

As is shown in Appendix \ref{PSF_effect}, the impact of the point spread function on the conclusions drawn from our sample is small. It contributes less than $0.005$ dex to the difference in N/O - O/H relation between extended and compact galaxies, and below $0.01$ dex to the difference in the local relationship between $\Sigma_{*}$ and log(N/O). The small size of this contribution is due to our strict constraints on the apparent size and inclination of galaxies in our sample.
The relationship between the resolved N/O distribution and the properties of galaxies on local and global scales is therefore complicated. Part of this complexity can be attributed to the differing production timescales of nitrogen and oxygen, and the mixing of metals cannot be ruled out as a mechanism for driving the N/O distribution away from a purely local scaling relation. The use of N/O as a diagnostic for galaxy evolution, or for setting the abundance scaling of O/H, must take both local and global effects into account.

\acknowledgements

We would like to thank the anonymous referee for their extremely useful comments on this manuscript. Their contributions have led to substantial improvements in this paper.
CT acknowledges NSF CAREER Award AST- 1554877.
M. A. B. acknowledges support from NSF-1814682.
J. G. F-T gratefully acknowledges the grant support provided by Proyecto Fondecyt Iniciaci\'on No. 11220340, and also from ANID Concurso de Fomento a la Vinculaci\'on Internacional para Instituciones de Investigaci\'on Regionales (Modalidad corta duraci\'on) Proyecto No. FOVI210020, and from the Joint Committee ESO-Government of Chile 2021 (ORP 023/2021). 

This research made use of \texttt{Astropy}, a community-developed core \texttt{python} package for astronomy \citep{Astropy2013,Astropy2018} and \texttt{matplotlib} \citep{Matplotlib}, an open-source \texttt{python} plotting library.

Funding for the Sloan Digital Sky 
Survey IV has been provided by the 
Alfred P. Sloan Foundation, the U.S. 
Department of Energy Office of 
Science, and the Participating 
Institutions. 

SDSS-IV acknowledges support and 
resources from the Center for High 
Performance Computing  at the 
University of Utah. The SDSS 
website is www.sdss.org.

SDSS-IV is managed by the 
Astrophysical Research Consortium 
for the Participating Institutions 
of the SDSS Collaboration including 
the Brazilian Participation Group, 
the Carnegie Institution for Science, 
Carnegie Mellon University, Center for 
Astrophysics | Harvard \& 
Smithsonian, the Chilean Participation 
Group, the French Participation Group, 
Instituto de Astrof\'isica de 
Canarias, The Johns Hopkins 
University, Kavli Institute for the 
Physics and Mathematics of the 
Universe (IPMU) / University of 
Tokyo, the Korean Participation Group, 
Lawrence Berkeley National Laboratory, 
Leibniz Institut f\"ur Astrophysik 
Potsdam (AIP),  Max-Planck-Institut 
f\"ur Astronomie (MPIA Heidelberg), 
Max-Planck-Institut f\"ur 
Astrophysik (MPA Garching), 
Max-Planck-Institut f\"ur 
Extraterrestrische Physik (MPE), 
National Astronomical Observatories of 
China, New Mexico State University, 
New York University, University of 
Notre Dame, Observat\'ario 
Nacional / MCTI, The Ohio State 
University, Pennsylvania State 
University, Shanghai 
Astronomical Observatory, United 
Kingdom Participation Group, 
Universidad Nacional Aut\'onoma 
de M\'exico, University of Arizona, 
University of Colorado Boulder, 
University of Oxford, University of 
Portsmouth, University of Utah, 
University of Virginia, University 
of Washington, University of 
Wisconsin, Vanderbilt University, 
and Yale University.

\appendix

\begin{appendices}

\section{Different indicators}\label{Different_indicators}

Studies of gas-phase chemical abundances in ionized gas with optical strong lines are plagued by the inherent uncertainties in their methodology. The difficulties of separating real changes in the chemical abundance from the variation of other conditions in the gas has been well studied \citep[see e.g.][]{Kewley2019}, but no strong line abundance indicator is without its drawbacks. In this Appendix we investigate the impact of other N/O diagnostics on our result. For this purpose we make use of the N/O calibration of \cite{Pilyugin2016}, and the N2O2 and N2S2 calibrations of \cite{PerezMontero2009}. The emission lines used by these estimators are listed in Table \ref{tab:indicators}. We will not repeat our full analysis, but provide copies of our key results derived with these alternative methods.

\begin{table*}[]
    \centering
    \begin{tabular}{|c|c|c|c|}
    \hline
    \multicolumn{4}{|c|}{\textbf{12+log(O/H)}} \\
    \hline
      Name & Lines & Notes & Reference\\
      \hline
    KK04 & R23\footnote{R23=([OII]$\lambda 3726,29$+[OIII]$4959,5007$)}+O32\footnote{O32 = [OIII]$\lambda 4959,5007$/[OII]$\lambda 3726,29$} & No dependence on ionisation parameter& \cite{Kobulnicky2004}\\
    \hline
    \multicolumn{4}{|c|}{\textbf{log(N/O)}} \\
    \hline
    T96& N2O2\footnote{N2O2=[NII]$\lambda 6584$/[OII]$\lambda 3726,29$} + R23 & & \cite{Thurston1996} \\
    PM\&C09 N2O2&N2O2&Empirically calibrated&\cite{PerezMontero2009} \\
    PM\&C09 N2S2&N2S2&Empirically calibrated&\cite{PerezMontero2009}\\
    PG16 N/O &N2$_{\beta}$\footnote{N2$_{\beta}$=[NII]6548,84/H$\beta$}, R2\footnote{R2=[OII]$\lambda 3726,29$/H$\beta$} & Empirically calibrated& \cite{Pilyugin2016}\\    
    \hline
    \end{tabular}
    \caption{A summary of the O/H and N/O indicators used for this work.} 
    \label{tab:indicators}
\end{table*}

\begin{figure*}
    \centering
    \includegraphics{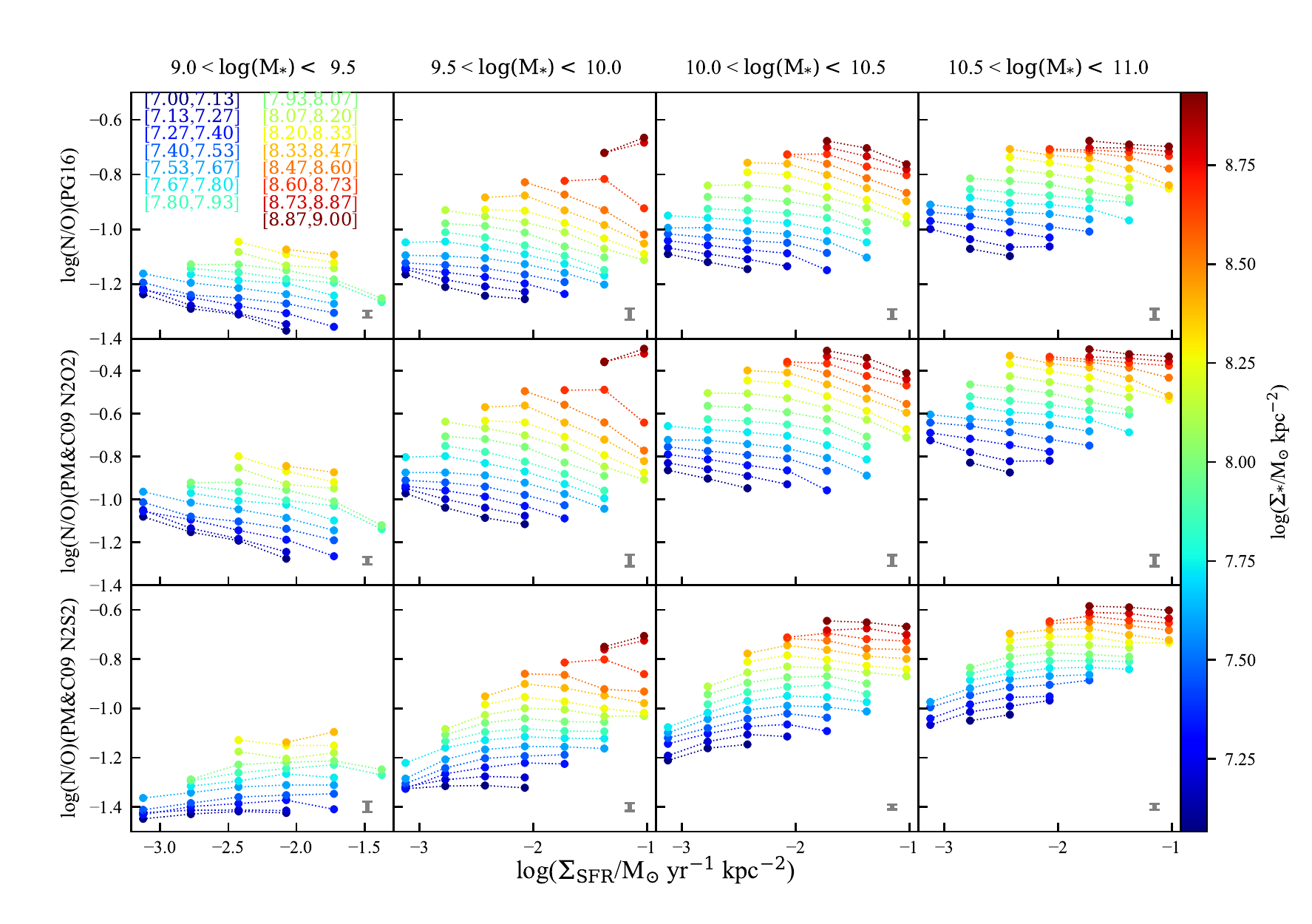}
    \caption{The local relationship between log(N/O), derived with three different N/O abundance calibrations, and $\log(\mathrm{\Sigma_{*}})$, $\log(\mathrm{\Sigma_{SFR}})$ and $\log(\mathrm{M_{*}})$. The top row shows the relationship for the \cite{Pilyugin2016} indicator, the second row shows the relationship for the \cite{PerezMontero2009} N2O2 indicator, while the bottom row displays the result for the \cite{PerezMontero2009} N2S2 indicator. The colour of each line represents the local $\log(\mathrm{\Sigma_{*}})$ in bins shown be the colour bar and the coloured intervals in the top left panel. The grey errorbars at the lower right of each panel indicate the largest bootstrapped uncertainty on any median shown therein.}
    \label{A1_FMNOR}
\end{figure*}

\begin{figure}
    \centering
    \includegraphics{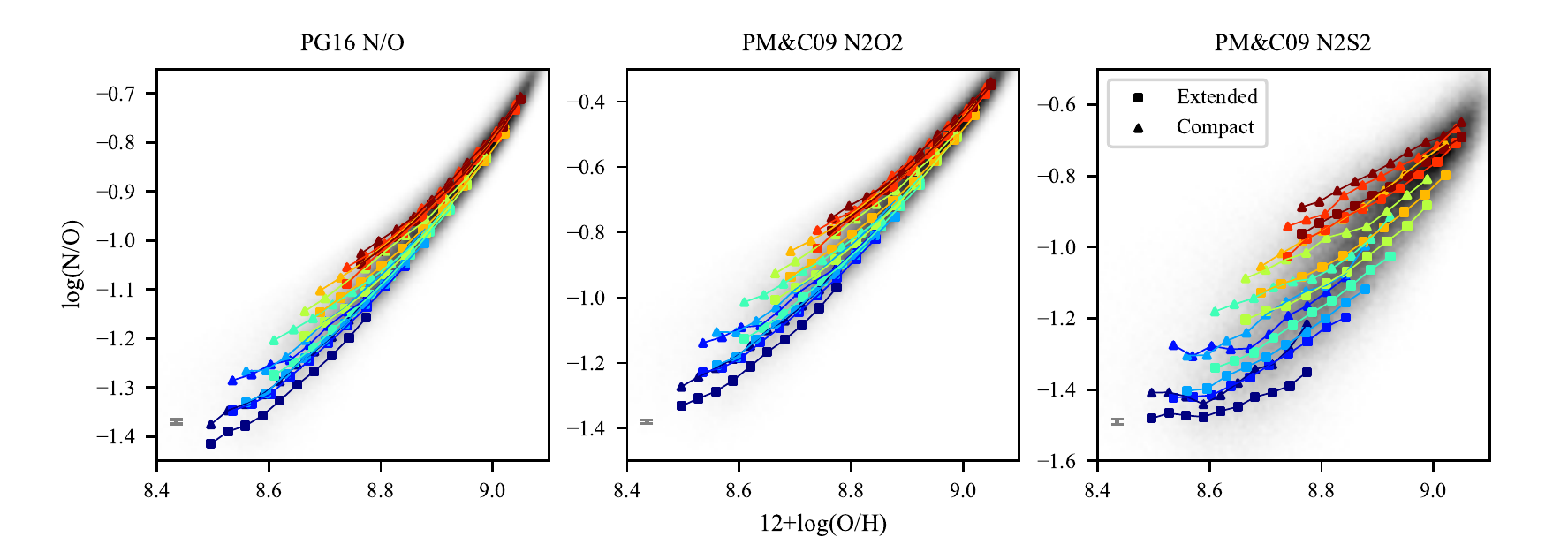}
    \caption{The relationship between N/O and O/H for three different N/O abundance diagnostics. The symbols are the same as in Figure \ref{NO_OH_comp_ext}, with squares representing extended galaxies and triangles denoting compact galaxies. Blue colours correspond to low-mass galaxies and red colours represent high-mass galaxies. The grey errorbars in the lower left of each panel are the largest bootstrapped uncertainties on any median calculated in the corresponding panel. In each case, 12+log(O/H) is estimated using the \cite{Kobulnicky2004} indicator. While each N/O indicator gives a result that is quantitatively different to what is seen in Figure \ref{NO_OH_comp_ext}, the total stellar mass trend remains, and the compact galaxies are always nitrogen-enhanced relative to the extended galaxies at a given O/H. }
    \label{A1_NO_OH}
\end{figure}

In Figure \ref{A1_FMNOR}, we show the relationship between N/O, $\mathrm{\Sigma_{*}}$, $\mathrm{\Sigma_{SFR}}$ in bins of total stellar mass as in Figure \ref{FMNOR_split_mass}. These three indicators show the same behaviour with total $\log(\mathrm{M_{*}})$ and local $\mathrm{\Sigma_{*}}$. While the \cite{Pilyugin2016} indicator and \cite{PerezMontero2009} N2O2 calibrations show the same reduction in N/O with increasing $\mathrm{\Sigma_{SFR}}$, the N2S2 calibration does not. In fact with the N2S2 calibration we see that the estimated log(N/O) increases with $\log(\mathrm{\Sigma_{SFR}})$. This difference in behaviour is likely attributable to a correlation between $\mathrm{\Sigma_{SFR}}$ and the ionisation parameter \citep{Mingozzi2020}. Since the ionization potential of $\mathrm{S^{+}}$ is $10.36$ eV while the ionization potential of $\mathrm{N^{+}}$ is $14.53$ eV, the [NII]/[SII] ratio will decline with increasing ionization potential, leading to a spurious apparent increase in N/O measured using this line ratio. This variation with ionisation parameter is less of an issue for indicators involving the [NII]/[OII] ratio, as the ionisation potential of $\mathrm{O^{+}}$ is closer to that of $\mathrm{N^{+}}$ at $13.61$ eV. We therefore regard the conclusion that N/O declines at higher $\mathrm{\Sigma_{SFR}}$ to be robust.

In Figure \ref{A1_NO_OH} we reproduce the N/O-O/H relation split by total stellar mass, for the compact and extended subsamples as initially shown in Figure \ref{NO_OH_comp_ext}. For all three alternative N/O indicators we see a clear and significant offset in N/O at fixed O/H between the compact and extended galaxies, with compact galaxies being more nitrogen-rich. Although the total range and scatter about the relation varies substantially between the different indicators, the main trends with stellar mass and size are broadly reproduced.

\section{The effect of the PSF on local scaling relations}\label{PSF_effect}
During the process of constructing the final data cubes, the MaNGA Data Reduction Pipeline smooths the data such that they have a $2\farcs5$ point spread function. While this produces a dataset with a uniform spatial resolution, this can have the effect of flattening gradients in the emission line intensities. Given that the abundance ratios are obtained through complicated non-linear combinations of emission line intensities, it is not obvious what effect this will have on the results that we have reported. The magnitudes of the changes in the measured quantities are related to the steepness of the light gradients in our observed galaxies. This means that the differences between the scaling relations in extended and compact galaxies may be induced by the PSF rather than being intrinsic to the galaxies.
Despite our selection being designed to minimise the effects of spatial resolution on our data \citep[recall that, following][$R_{e}>4\arcsec$ was a criterion for inclusion in our sample]{Belfiore2017}, it will be useful to quantify the effect of the PSF on our results.

\begin{figure}
    \centering
    \includegraphics{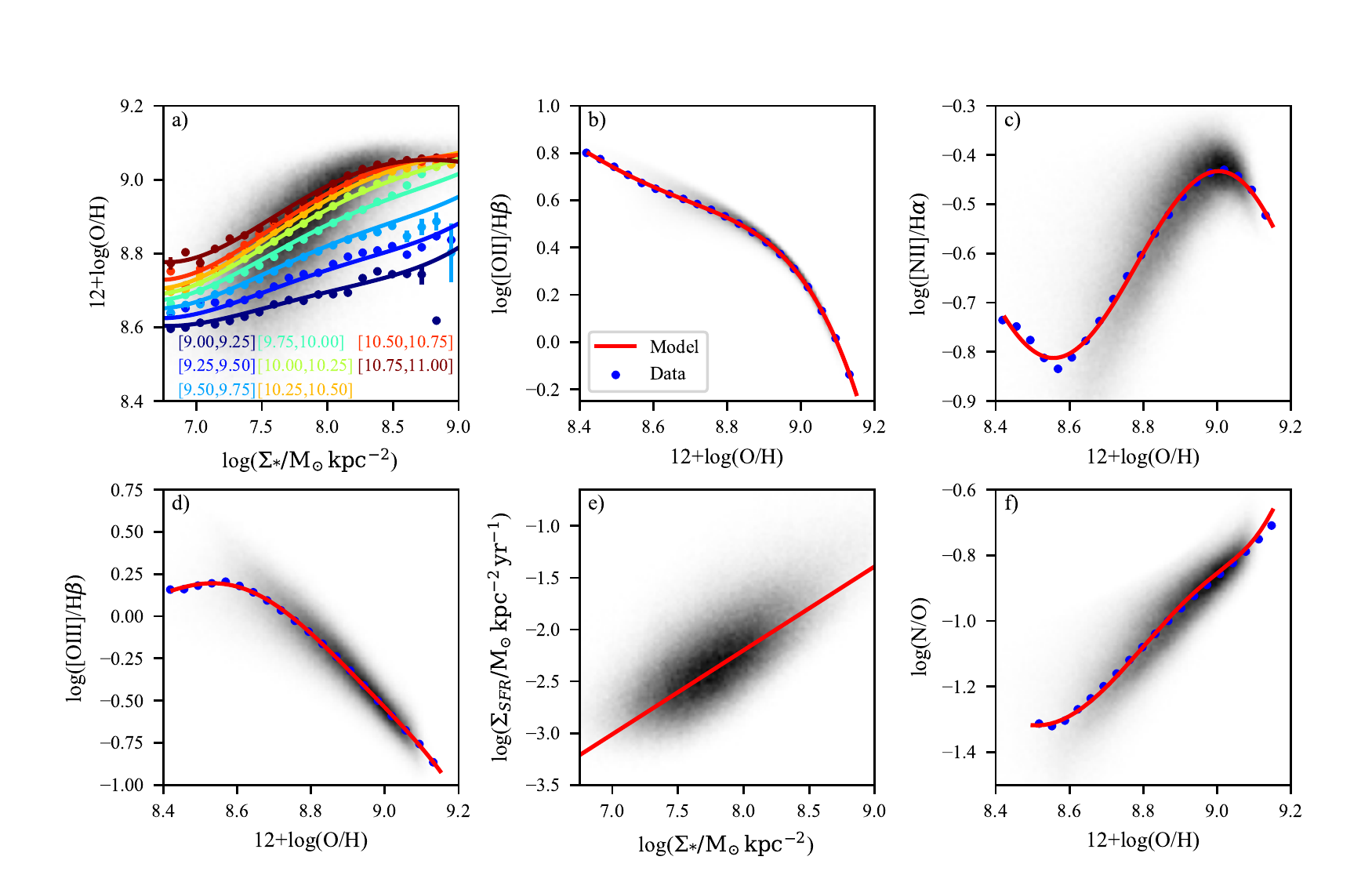}
    \caption{The relations between observables for MaNGA galaxies in our sample, and the fits to these relations that are used to build the models. In each panel, the dotted data points are medians of MaNGA spaxel data in narrow bins of the quantity shown on the x-axis of each panel. In each panel, the grey represents the density of observed data points for the entire sample. \emph{a)} The total-stellar-mass dependent local relationship between $\Sigma_{*}$ and 12+log(O/H). Coloured points are the median 12+log(O/H) in bins of $\log(\Sigma_{*})$ and $\log(M_{*})$, while coloured lines are the estimated oxygen abundances from our two-dimensional polynomial model. The colors of each line correspond to the intervals of total $\log(M_{*})$ shown in brackets of the same colour at the bottom of the panel.\emph{b-d)} The dust-corrected emission line ratios as a function of oxygen abundance. Blue points are the median ratio value in narrow bins of 12+log(O/H), and the red lines are 5th degree polynomial fits to the medians. \emph{e)} The relationship between local $\Sigma_{*}$ and $\Sigma_{SFR}$. \emph{f)} Blue points show the median N/O-O/H relation from MaNGA data, and the red line shows the estimated N/O derived from inputting the modelled line ratios into Equation \ref{NO_thurston}. }
    \label{poly_models}
\end{figure}

To do so, we have created a set of model galaxies which were then investigated with and without the effect of an instrumental PSF. The fundamental component of these models is an axisymmetric exponential disk with total stellar mass $\mathrm{M_{*}}$ and a half-light radius of $R_{e}$. The stellar mass surface density is described by $\Sigma_{*}(R)=a10^{-bR}$, where $a$ is the central stellar surface density and $b$ describes how sharply the stellar density drops with radius, $R$. In order to investigate the impact of the PSF on the observed abundance distributions, we take an empirical approach to imbuing our modelled galaxies with the appropriate emission line fluxes. We fit the observed relationship between $\log(\Sigma_{*})$, $\log( \mathrm{M_{*}})$ and $12+\mathrm{\log(O/H)}$ with a general two dimensional polynomial of the form $\sum\limits_{i,j}c_{i,j}x^{i}y^{j}$ with degree $4$ over the ranges $6.75< \log(\Sigma_{*})<9.0$ and $9<\log{\mathrm{M_{*}}}<11$. This polynomial is used to assign a gas-phase metallicity to each point in the model galaxies based on the $\mathrm{M_{*}}$ and $\Sigma_{*}$. To assign a line flux at each location in the galaxy, we fit the observed relationship between $12+\mathrm{\log(O/H)}$ and the dust-corrected log([OIII]5007/H$\beta$), log([OII]3726,3729/H$\beta$) and log([NII]6584/H$\alpha$) with a $5$th degree polynomial.
Finally, we scale these line ratios by the local star formation rate surface density, which is linearly proportional to the Balmer line flux. $\Sigma_{SFR}$ is estimated from the fitted $\Sigma_{SFR} - \Sigma_{*}$ relation derived in section \ref{SFRD}. Each model galaxy is created with a resolution of $500\times 500$ pixels. The $\Sigma_{*}$ and line intensity maps are then convolved with the PSF before the abundance maps are produced. Finally, these maps are resampled onto a grid of $0\farcs 5$ to match the MaNGA data. The relations between $\Sigma_{*}$, the emission line ratios, metallicities and $\Sigma_{SFR}$ are shown in Figure \ref{poly_models}.

We note that this prescription for assigning [NII]6584 fluxes based on the oxygen abundance tacitly assumes a universal relationship between N/O and O/H. The nitrogen abundance excess in the outer parts of galaxies that we have observed imply that the [NII]6584 gradients are in fact shallower than our models suggest. This would reduce any changes to the N/O gradient imposed by the PSF.

\begin{figure}
    \centering
    \includegraphics{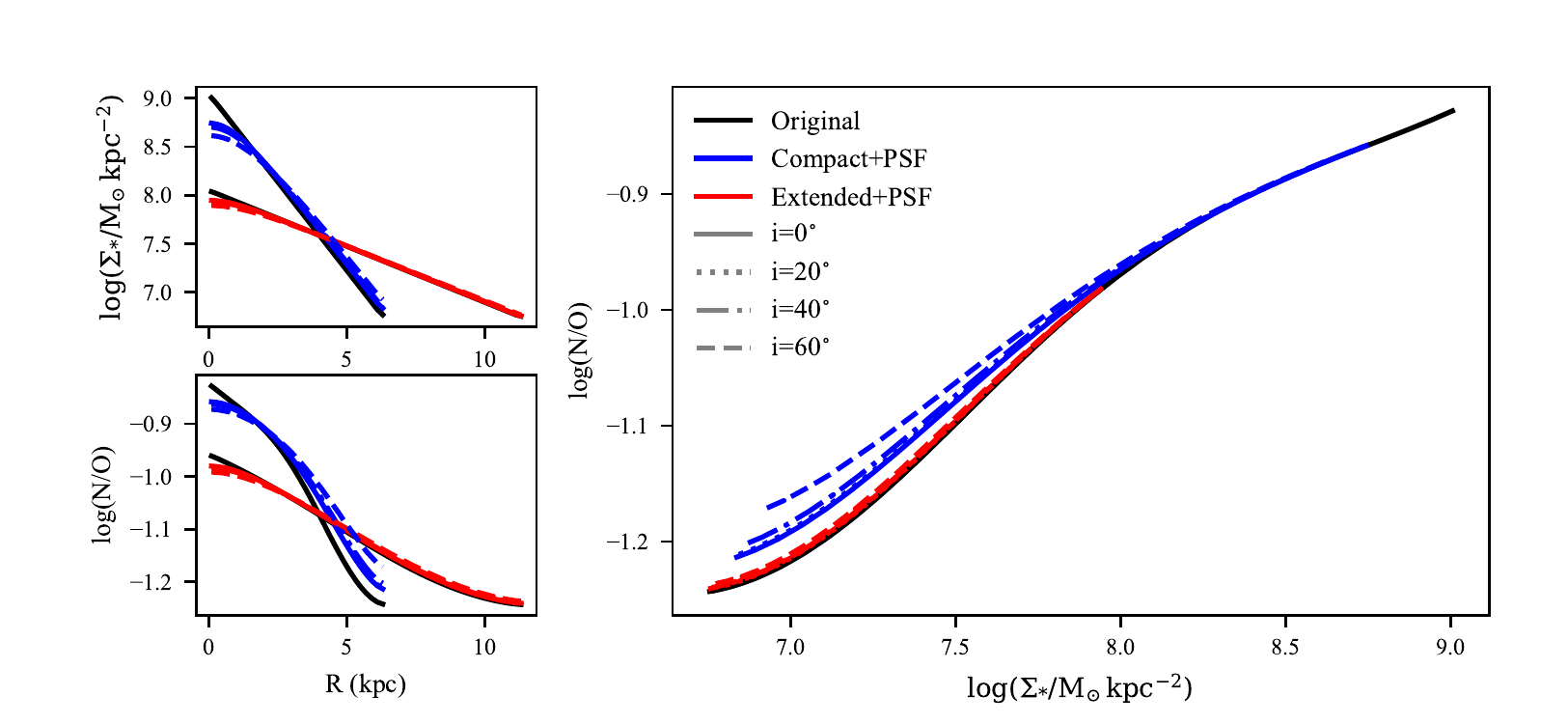}
    \caption{The effect of compactness and inclination on the observed distributions of $\Sigma_{*}$ and log(N/O) in galaxies. We have presented the radial profiles of $\Sigma_{*}$ and log(N/O) as well as the $\Sigma_{*}$-log(N/O) relation for two galaxies with $\log(M_{*})=10.5$. $\mathrm{R}-\Sigma_{*}$ is shown at the upper left, while $\mathrm{R - \log N/O}$ is shown in the lower left. The large panel on the right shows the $\Sigma_{*}$-log(N/O) relation. Solid black lines show the original models, with red and blue lines indicating the extended and compact galaxies respectively. Different line styles correspond to different inclinations as indicated by the legend in the right hand panel. The effect of the PSF is most apparent in the highly inclined and compact models, though the most extreme case shifts the log(N/O) by no more than $0.05$ dex. }
    \label{single_galaxy_w_psf}
\end{figure}

With these models we tested the effect of convolution of the mass and line intensity maps with the PSF on the local scaling relations. We find that convolution with a Gaussian PSF flattens the radial gradients in all emission line intensities and abundance ratios, as well as the stellar mass density map. The degree to which the gradients change depends on the apparent size of the galaxy relative to the PSF and the concentration of light, which is determined by the half light radius and redshift in our models. Gradients of quantities are further effected by the inclination of the galaxy relative to the line of sight, with highly inclined galaxies suffering from beam smearing more than face-on galaxies. We demonstrate this effect in Figure \ref{single_galaxy_w_psf}, where we show the radial profiles of $\Sigma_{*}$ and log(N/O) for two galaxies with $\log(\mathrm{M_{\odot}})=10.5$ at a redshift of  $z=0.035$, before and after convolution with a $2\farcs5$ PSF. We generate a compact galaxy with a half-mass radius of $10^{0.3}=2.0 \, \mathrm{kpc}$ and an extended galaxy with a half-mass radius of $10^{0.8}=6.3 \, \mathrm{kpc}$. These are very extreme sizes for galaxies of this mass in our sample, and thus provide upper and lower limits on the extent of beam smearing on galaxies in our sample. The galaxy models are generated with four inclinations between face-on ($0^{\circ}$) and $60^{\circ}$, which is slightly higher than the highest inclination of galaxies in our sample.

The effect of the PSF is to reduce both $\Sigma_{*}$ and log(N/O) in the centre of the galaxies and to increase these quantities in the outer parts. The change in these quantities is higher in the more compact galaxy but is more pronounced for log(N/O). This means that the log(N/O) increases at fixed $\Sigma_{*}$. The greatest possible change in log(N/O) occurs at low stellar mass surface density, but is never larger than $0.05$ dex in the most extremely compact and inclined system. While this effect is large enough to be detected, we note that the vast majority of galaxies in our sample are less inclined.

\begin{figure}
    \centering
    \includegraphics{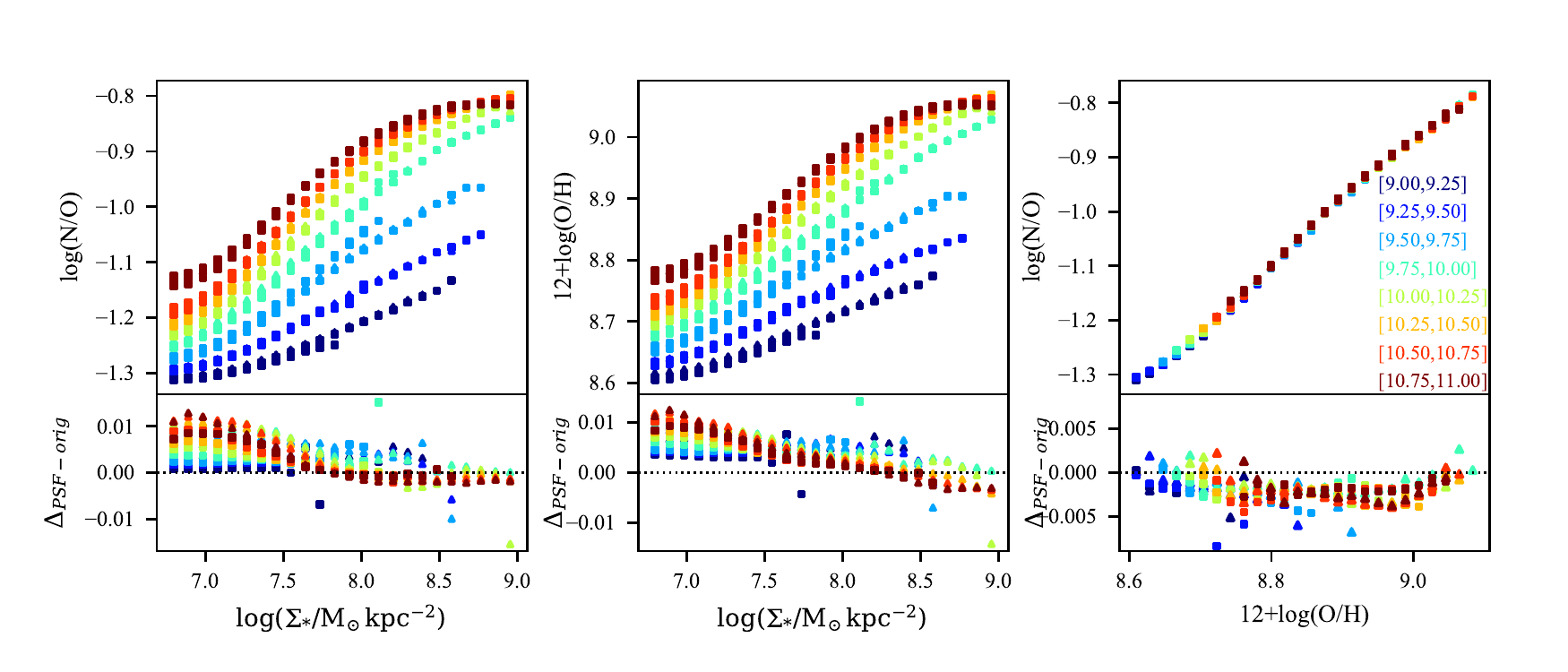}
    \caption{The effect of the PSF on the local scaling relations in narrow bins of total stellar mass for a simulated set of galaxies matching the MaNGA sample in mass, size, redshift and inclination. Panels in the top row show the log(N/O) and log(O/H) as a function of $\Sigma_{*}$ for both compact (triangles) and extended (squares) galaxies. The colors of the points correspond to an interval of total stellar mass indicated by text of the corresponding color in the right hand panel. In the lower row we show the difference in log(N/O) or log(O/H) at a given $\log(\Sigma_{*})$ imposed by the PSF. For our sample, the PSF explains at most $\sim 0.01$ dex in the difference between the compact and extended subsamples. The greatest difference occurs in galaxies with $\log(\mathrm{M_{*}})\sim 10.5$ with $\log(\Sigma_{*})<7.5$, and is smaller than $0.005$ for galaxies in the lowest stellar mass bins. In the right hand panes we show the effect of the PSF on the N/O-O/H relation. The difference in N/O imposed by the PSF at fixed O/H is universally smaller than $0.005$ dex, regardless of galaxy size.}
    \label{PSF_adjusted_scaling_relations}
\end{figure}

Our sample includes galaxies with a range of physical sizes, redshifts and inclinations. To test the total effect of the PSF on the results reported in the body of this paper we generate a simulated sample of galaxies that matches the stellar masses, sizes, inclinations and redshifts of the sample described in Section \ref{Galaxy_selection}. We then produce the local $\Sigma_{*}$ - N/O, $\Sigma_{*}$-O/H and N/O-O/H relations before and after the effect of the PSF is applied to the data. We repeat this process for the simulated compact and extended subsamples. The results of this test are shown in Figure \ref{PSF_adjusted_scaling_relations}. 

The change in the local $\Sigma_{*}$-N/O and $\Sigma_{*}$-O/H scaling relations induced by the PSF is largest for the compact subsample of galaxies. The difference is largest at low stellar mass surface densities, corresponding to the outer regions of galaxies. Our models suggest that for our sample of galaxies, the change in these relations is at most $0.01$ dex in both N/O and O/H in the compact subsample. In the extended galaxies, the change in scaling relations is even smaller. In combination, the changes in the $\Sigma_{*}$-N/O and O/H relations do not affect the N/O-O/H relation at a level greater than $0.005$ dex. Not only do we rule out the role of the PSF in setting up the observed differences between compact and extended galaxies, but we can also rule out the role of the PSF in inducing the total stellar mass-dependence of the N/O-O/H relation. The main conclusions of this paper are not significantly affected by the spatial resolution of the MaNGA data. The only way to explain the differences in the samples that we see is for the chemical abundance properties of galaxies to differ between the extended and compact subsamples.

\end{appendices}

\bibliography{bibliography}

\begin{thebibliography}{}
\expandafter\ifx\csname natexlab\endcsname\relax\def\natexlab#1{#1}\fi
\providecommand{\url}[1]{\href{#1}{#1}}
\providecommand{\dodoi}[1]{doi:~\href{http://doi.org/#1}{\nolinkurl{#1}}}
\providecommand{\doeprint}[1]{\href{http://ascl.net/#1}{\nolinkurl{http://ascl.net/#1}}}
\providecommand{\doarXiv}[1]{\href{https://arxiv.org/abs/#1}{\nolinkurl{https://arxiv.org/abs/#1}}}

\bibitem[{{Andrews} \& {Martini}(2013)}]{Andrews2013}
{Andrews}, B.~H., \& {Martini}, P. 2013, \apj, 765, 140,
  \dodoi{10.1088/0004-637X/765/2/140}

\bibitem[{{Astropy Collaboration} {et~al.}(2013){Astropy Collaboration},
  {Robitaille}, {Tollerud}, {Greenfield}, {Droettboom}, {Bray}, {Aldcroft},
  {Davis}, {Ginsburg}, {Price-Whelan}, {Kerzendorf}, {Conley}, {Crighton},
  {Barbary}, {Muna}, {Ferguson}, {Grollier}, {Parikh}, {Nair}, {Unther},
  {Deil}, {Woillez}, {Conseil}, {Kramer}, {Turner}, {Singer}, {Fox}, {Weaver},
  {Zabalza}, {Edwards}, {Azalee Bostroem}, {Burke}, {Casey}, {Crawford},
  {Dencheva}, {Ely}, {Jenness}, {Labrie}, {Lim}, {Pierfederici}, {Pontzen},
  {Ptak}, {Refsdal}, {Servillat}, \& {Streicher}}]{Astropy2013}
{Astropy Collaboration}, {Robitaille}, T.~P., {Tollerud}, E.~J., {et~al.} 2013,
  \aap, 558, A33, \dodoi{10.1051/0004-6361/201322068}

\bibitem[{{Astropy Collaboration} {et~al.}(2018){Astropy Collaboration},
  {Price-Whelan}, {Sip{\H o}cz}, {G{\"u}nther}, {Lim}, {Crawford}, {Conseil},
  {Shupe}, {Craig}, {Dencheva}, {Ginsburg}, {VanderPlas}, {Bradley},
  {P{\'e}rez-Su{\'a}rez}, {de Val-Borro}, {Aldcroft}, {Cruz}, {Robitaille},
  {Tollerud}, {Ardelean}, {Babej}, {Bach}, {Bachetti}, {Bakanov}, {Bamford},
  {Barentsen}, {Barmby}, {Baumbach}, {Berry}, {Biscani}, {Boquien}, {Bostroem},
  {Bouma}, {Brammer}, {Bray}, {Breytenbach}, {Buddelmeijer}, {Burke},
  {Calderone}, {Cano Rodr{\'{\i}}guez}, {Cara}, {Cardoso}, {Cheedella},
  {Copin}, {Corrales}, {Crichton}, {D'Avella}, {Deil}, {Depagne}, {Dietrich},
  {Donath}, {Droettboom}, {Earl}, {Erben}, {Fabbro}, {Ferreira}, {Finethy},
  {Fox}, {Garrison}, {Gibbons}, {Goldstein}, {Gommers}, {Greco}, {Greenfield},
  {Groener}, {Grollier}, {Hagen}, {Hirst}, {Homeier}, {Horton}, {Hosseinzadeh},
  {Hu}, {Hunkeler}, {Ivezi{\'c}}, {Jain}, {Jenness}, {Kanarek}, {Kendrew},
  {Kern}, {Kerzendorf}, {Khvalko}, {King}, {Kirkby}, {Kulkarni}, {Kumar},
  {Lee}, {Lenz}, {Littlefair}, {Ma}, {Macleod}, {Mastropietro}, {McCully},
  {Montagnac}, {Morris}, {Mueller}, {Mumford}, {Muna}, {Murphy}, {Nelson},
  {Nguyen}, {Ninan}, {N{\"o}the}, {Ogaz}, {Oh}, {Parejko}, {Parley}, {Pascual},
  {Patil}, {Patil}, {Plunkett}, {Prochaska}, {Rastogi}, {Reddy Janga},
  {Sabater}, {Sakurikar}, {Seifert}, {Sherbert}, {Sherwood-Taylor}, {Shih},
  {Sick}, {Silbiger}, {Singanamalla}, {Singer}, {Sladen}, {Sooley},
  {Sornarajah}, {Streicher}, {Teuben}, {Thomas}, {Tremblay}, {Turner},
  {Terr{\'o}n}, {van Kerkwijk}, {de la Vega}, {Watkins}, {Weaver}, {Whitmore},
  {Woillez}, {Zabalza}, \& {Astropy Contributors}}]{Astropy2018}
{Astropy Collaboration}, {Price-Whelan}, A.~M., {Sip{\H o}cz}, B.~M., {et~al.}
  2018, \aj, 156, 123, \dodoi{10.3847/1538-3881/aabc4f}

\bibitem[{{Balogh} {et~al.}(1999){Balogh}, {Morris}, {Yee}, {Carlberg}, \&
  {Ellingson}}]{Balogh1999}
{Balogh}, M.~L., {Morris}, S.~L., {Yee}, H.~K.~C., {Carlberg}, R.~G., \&
  {Ellingson}, E. 1999, \apj, 527, 54, \dodoi{10.1086/308056}

\bibitem[{{Barrera-Ballesteros} {et~al.}(2016){Barrera-Ballesteros}, {Heckman},
  {Zhu}, {Zakamska}, {S{\'a}nchez}, {Law}, {Wake}, {Green}, {Bizyaev},
  {Oravetz}, {Simmons}, {Malanushenko}, {Pan}, {Roman Lopes}, \&
  {Lane}}]{BarreraBallesteros2016}
{Barrera-Ballesteros}, J.~K., {Heckman}, T.~M., {Zhu}, G.~B., {et~al.} 2016,
  \mnras, 463, 2513, \dodoi{10.1093/mnras/stw1984}

\bibitem[{{Barrera-Ballesteros} {et~al.}(2018){Barrera-Ballesteros}, {Heckman},
  {S{\'a}nchez}, {Zakamska}, {Cleary}, {Zhu}, {Brinkmann}, {Drory}, \& {THE
  MaNGA TEAM}}]{BarreraBallesteros2018}
{Barrera-Ballesteros}, J.~K., {Heckman}, T., {S{\'a}nchez}, S.~F., {et~al.}
  2018, \apj, 852, 74, \dodoi{10.3847/1538-4357/aa9b31}

\bibitem[{{Belfiore} {et~al.}(2015){Belfiore}, {Maiolino}, {Bundy}, {Thomas},
  {Maraston}, {Wilkinson}, {S{\'a}nchez}, {Bershady}, {Blanc}, {Bothwell},
  {Cales}, {Coccato}, {Drory}, {Emsellem}, {Fu}, {Gelfand}, {Law}, {Masters},
  {Parejko}, {Tremonti}, {Wake}, {Weijmans}, {Yan}, {Xiao}, {Zhang}, {Zheng},
  {Bizyaev}, {Kinemuchi}, {Oravetz}, \& {Simmons}}]{Belfiore2015}
{Belfiore}, F., {Maiolino}, R., {Bundy}, K., {et~al.} 2015, \mnras, 449, 867,
  \dodoi{10.1093/mnras/stv296}

\bibitem[{{Belfiore} {et~al.}(2017){Belfiore}, {Maiolino}, {Tremonti},
  {S{\'a}nchez}, {Bundy}, {Bershady}, {Westfall}, {Lin}, {Drory}, {Boquien},
  {Thomas}, \& {Brinkmann}}]{Belfiore2017}
{Belfiore}, F., {Maiolino}, R., {Tremonti}, C., {et~al.} 2017, \mnras, 469,
  151, \dodoi{10.1093/mnras/stx789}

\bibitem[{{Belfiore} {et~al.}(2019){Belfiore}, {Westfall}, {Schaefer},
  {Cappellari}, {Ji}, {Bershady}, {Tremonti}, {Law}, {Yan}, {Bundy}, {Shetty},
  {Drory}, {Thomas}, {Emsellem}, \& {S{\'a}nchez}}]{Belfiore2019}
{Belfiore}, F., {Westfall}, K.~B., {Schaefer}, A., {et~al.} 2019, arXiv
  e-prints.
\newblock \doarXiv{1901.00866}

\bibitem[{{Bian} {et~al.}(2020){Bian}, {Kewley}, {Groves}, \&
  {Dopita}}]{Bian2020}
{Bian}, F., {Kewley}, L.~J., {Groves}, B., \& {Dopita}, M.~A. 2020, \mnras,
  493, 580, \dodoi{10.1093/mnras/staa259}

\bibitem[{{Blanton} {et~al.}(2017){Blanton}, {Bershady}, {Abolfathi},
  {Albareti}, {Allende Prieto}, {Almeida}, {Alonso-Garc{\'\i}a}, {Anders},
  {Anderson}, {Andrews}, {Aquino-Ort{\'\i}z}, {Arag{\'o}n-Salamanca},
  {Argudo-Fern{\'a}ndez}, {Armengaud}, {Aubourg}, {Avila-Reese}, {Badenes},
  {Bailey}, {Barger}, {Barrera-Ballesteros}, {Bartosz}, {Bates}, {Baumgarten},
  {Bautista}, {Beaton}, {Beers}, {Belfiore}, {Bender}, {Berlind}, {Bernardi},
  {Beutler}, {Bird}, {Bizyaev}, {Blanc}, {Blomqvist}, {Bolton}, {Boquien},
  {Borissova}, {van den Bosch}, {Bovy}, {Brandt}, {Brinkmann}, {Brownstein},
  {Bundy}, {Burgasser}, {Burtin}, {Busca}, {Cappellari}, {Delgado Carigi},
  {Carlberg}, {Carnero Rosell}, {Carrera}, {Chanover}, {Cherinka}, {Cheung},
  {G{\'o}mez Maqueo Chew}, {Chiappini}, {Choi}, {Chojnowski}, {Chuang},
  {Chung}, {Cirolini}, {Clerc}, {Cohen}, {Comparat}, {da Costa}, {Cousinou},
  {Covey}, {Crane}, {Croft}, {Cruz-Gonzalez}, {Garrido Cuadra}, {Cunha},
  {Damke}, {Darling}, {Davies}, {Dawson}, {de la Macorra}, {Dell'Agli}, {De
  Lee}, {Delubac}, {Di Mille}, {Diamond-Stanic}, {Cano-D{\'\i}az}, {Donor},
  {Downes}, {Drory}, {du Mas des Bourboux}, {Duckworth}, {Dwelly}, {Dyer},
  {Ebelke}, {Eigenbrot}, {Eisenstein}, {Emsellem}, {Eracleous}, {Escoffier},
  {Evans}, {Fan}, {Fern{\'a}ndez-Alvar}, {Fernandez-Trincado}, {Feuillet},
  {Finoguenov}, {Fleming}, {Font-Ribera}, {Fredrickson}, {Freischlad},
  {Frinchaboy}, {Fuentes}, {Galbany}, {Garcia-Dias},
  {Garc{\'\i}a-Hern{\'a}ndez}, {Gaulme}, {Geisler}, {Gelfand},
  {Gil-Mar{\'\i}n}, {Gillespie}, {Goddard}, {Gonzalez-Perez}, {Grabowski},
  {Green}, {Grier}, {Gunn}, {Guo}, {Guy}, {Hagen}, {Hahn}, {Hall}, {Harding},
  {Hasselquist}, {Hawley}, {Hearty}, {Gonzalez Hern{\'a}ndez}, {Ho}, {Hogg},
  {Holley-Bockelmann}, {Holtzman}, {Holzer}, {Huehnerhoff}, {Hutchinson},
  {Hwang}, {Ibarra-Medel}, {da Silva Ilha}, {Ivans}, {Ivory}, {Jackson},
  {Jensen}, {Johnson}, {Jones}, {J{\"o}nsson}, {Jullo}, {Kamble}, {Kinemuchi},
  {Kirkby}, {Kitaura}, {Klaene}, {Knapp}, {Kneib}, {Kollmeier}, {Lacerna},
  {Lane}, {Lang}, {Law}, {Lazarz}, {Lee}, {Le Goff}, {Liang}, {Li}, {Li},
  {Lian}, {Lima}, {Lin}, {Lin}, {Bertran de Lis}, {Liu}, {de Icaza Lizaola},
  {Long}, {Lucatello}, {Lundgren}, {MacDonald}, {Deconto Machado}, {MacLeod},
  {Mahadevan}, {Geimba Maia}, {Maiolino}, {Majewski}, {Malanushenko},
  {Malanushenko}, {Manchado}, {Mao}, {Maraston}, {Marques-Chaves}, {Masseron},
  {Masters}, {McBride}, {McDermid}, {McGrath}, {McGreer}, {Medina Pe{\~n}a},
  {Melendez}, {Merloni}, {Merrifield}, {Meszaros}, {Meza}, {Minchev},
  {Minniti}, {Miyaji}, {More}, {Mulchaey}, {M{\"u}ller-S{\'a}nchez}, {Muna},
  {Munoz}, {Myers}, {Nair}, {Nandra}, {Correa do Nascimento}, {Negrete},
  {Ness}, {Newman}, {Nichol}, {Nidever}, {Nitschelm}, {Ntelis}, {O'Connell},
  {Oelkers}, {Oravetz}, {Oravetz}, {Pace}, {Padilla}, {Palanque-Delabrouille},
  {Alonso Palicio}, {Pan}, {Parejko}, {Parikh}, {P{\^a}ris}, {Park}, {Patten},
  {Peirani}, {Pellejero-Ibanez}, {Penny}, {Percival}, {Perez-Fournon},
  {Petitjean}, {Pieri}, {Pinsonneault}, {Pisani}, {Poleski}, {Prada},
  {Prakash}, {Queiroz}, {Raddick}, {Raichoor}, {Barboza Rembold}, {Richstein},
  {Riffel}, {Riffel}, {Rix}, {Robin}, {Rockosi}, {Rodr{\'\i}guez-Torres},
  {Roman-Lopes}, {Rom{\'a}n-Z{\'u}{\~n}iga}, {Rosado}, {Ross}, {Rossi}, {Ruan},
  {Ruggeri}, {Rykoff}, {Salazar-Albornoz}, {Salvato}, {S{\'a}nchez}, {Aguado},
  {S{\'a}nchez-Gallego}, {Santana}, {Santiago}, {Sayres}, {Schiavon}, {da Silva
  Schimoia}, {Schlafly}, {Schlegel}, {Schneider}, {Schultheis}, {Schuster},
  {Schwope}, {Seo}, {Shao}, {Shen}, {Shetrone}, {Shull}, {Simon}, {Skinner},
  {Skrutskie}, {Slosar}, {Smith}, {Sobeck}, {Sobreira}, {Somers}, {Souto},
  {Stark}, {Stassun}, {Stauffer}, {Steinmetz}, {Storchi-Bergmann},
  {Streblyanska}, {Stringfellow}, {Su{\'a}rez}, {Sun}, {Suzuki}, {Szigeti},
  {Taghizadeh-Popp}, {Tang}, {Tao}, {Tayar}, {Tembe}, {Teske}, {Thakar},
  {Thomas}, {Thompson}, {Tinker}, {Tissera}, {Tojeiro}, {Hernandez Toledo}, {de
  la Torre}, {Tremonti}, {Troup}, {Valenzuela}, {Martinez Valpuesta},
  {Vargas-Gonz{\'a}lez}, {Vargas-Maga{\~n}a}, {Vazquez}, {Villanova}, {Vivek},
  {Vogt}, {Wake}, {Walterbos}, {Wang}, {Weaver}, {Weijmans}, {Weinberg},
  {Westfall}, {Whelan}, {Wild}, {Wilson}, {Wood-Vasey}, {Wylezalek}, {Xiao},
  {Yan}, {Yang}, {Ybarra}, {Y{\`e}che}, {Zakamska}, {Zamora}, {Zarrouk},
  {Zasowski}, {Zhang}, {Zhao}, {Zheng}, {Zheng}, {Zhou}, {Zhou}, {Zhu},
  {Zoccali}, \& {Zou}}]{Blanton2017}
{Blanton}, M.~R., {Bershady}, M.~A., {Abolfathi}, B., {et~al.} 2017, \aj, 154,
  28, \dodoi{10.3847/1538-3881/aa7567}

\bibitem[{{Bluck} {et~al.}(2020){Bluck}, {Maiolino}, {S{\'a}nchez}, {Ellison},
  {Thorp}, {Piotrowska}, {Teimoorinia}, \& {Bundy}}]{Bluck2020}
{Bluck}, A. F.~L., {Maiolino}, R., {S{\'a}nchez}, S.~F., {et~al.} 2020, \mnras,
  492, 96, \dodoi{10.1093/mnras/stz3264}

\bibitem[{{Boardman} {et~al.}(2021){Boardman}, {Zasowski}, {Newman}, {Sanchez},
  {Schaefer}, {Lian}, {Bizyaev}, \& {Drory}}]{Boardman2021}
{Boardman}, N.~F., {Zasowski}, G., {Newman}, J.~A., {et~al.} 2021, \mnras, 501,
  948, \dodoi{10.1093/mnras/staa3785}

\bibitem[{{Bregman}(1980)}]{Bregman1980}
{Bregman}, J.~N. 1980, \apj, 236, 577, \dodoi{10.1086/157776}

\bibitem[{{Bundy} {et~al.}(2015){Bundy}, {Bershady}, {Law}, {Yan}, {Drory},
  {MacDonald}, {Wake}, {Cherinka}, {S{\'a}nchez-Gallego}, {Weijmans}, {Thomas},
  {Tremonti}, {Masters}, {Coccato}, {Diamond-Stanic}, {Arag{\'o}n-Salamanca},
  {Avila-Reese}, {Badenes}, {Falc{\'o}n-Barroso}, {Belfiore}, {Bizyaev},
  {Blanc}, {Bland-Hawthorn}, {Blanton}, {Brownstein}, {Byler}, {Cappellari},
  {Conroy}, {Dutton}, {Emsellem}, {Etherington}, {Frinchaboy}, {Fu}, {Gunn},
  {Harding}, {Johnston}, {Kauffmann}, {Kinemuchi}, {Klaene}, {Knapen},
  {Leauthaud}, {Li}, {Lin}, {Maiolino}, {Malanushenko}, {Malanushenko}, {Mao},
  {Maraston}, {McDermid}, {Merrifield}, {Nichol}, {Oravetz}, {Pan}, {Parejko},
  {Sanchez}, {Schlegel}, {Simmons}, {Steele}, {Steinmetz}, {Thanjavur},
  {Thompson}, {Tinker}, {van den Bosch}, {Westfall}, {Wilkinson}, {Wright},
  {Xiao}, \& {Zhang}}]{Bundy2015}
{Bundy}, K., {Bershady}, M.~A., {Law}, D.~R., {et~al.} 2015, \apj, 798, 7,
  \dodoi{10.1088/0004-637X/798/1/7}

\bibitem[{{Burbidge} {et~al.}(1957){Burbidge}, {Burbidge}, {Fowler}, \&
  {Hoyle}}]{B2FH}
{Burbidge}, E.~M., {Burbidge}, G.~R., {Fowler}, W.~A., \& {Hoyle}, F. 1957,
  Reviews of Modern Physics, 29, 547, \dodoi{10.1103/RevModPhys.29.547}

\bibitem[{{Cappellari}(2017)}]{Cappellari2017}
{Cappellari}, M. 2017, \mnras, 466, 798, \dodoi{10.1093/mnras/stw3020}

\bibitem[{{Cappellari} \& {Emsellem}(2004)}]{Cappellari2004}
{Cappellari}, M., \& {Emsellem}, E. 2004, \pasp, 116, 138,
  \dodoi{10.1086/381875}

\bibitem[{{Chabrier}(2003)}]{Chabrier2003}
{Chabrier}, G. 2003, \pasp, 115, 763, \dodoi{10.1086/376392}

\bibitem[{{Cid Fernandes} {et~al.}(2010){Cid Fernandes}, {Stasi{\'n}ska},
  {Schlickmann}, {Mateus}, {Vale Asari}, {Schoenell}, \&
  {Sodr{\'e}}}]{CidFernandes2010}
{Cid Fernandes}, R., {Stasi{\'n}ska}, G., {Schlickmann}, M.~S., {et~al.} 2010,
  \mnras, 403, 1036, \dodoi{10.1111/j.1365-2966.2009.16185.x}

\bibitem[{{Cid Fernandes} {et~al.}(2013){Cid Fernandes}, {P{\'e}rez},
  {Garc{\'\i}a Benito}, {Gonz{\'a}lez Delgado}, {de Amorim}, {S{\'a}nchez},
  {Husemann}, {Falc{\'o}n Barroso}, {S{\'a}nchez-Bl{\'a}zquez}, {Walcher}, \&
  {Mast}}]{CidFernandes2013}
{Cid Fernandes}, R., {P{\'e}rez}, E., {Garc{\'\i}a Benito}, R., {et~al.} 2013,
  \aap, 557, A86, \dodoi{10.1051/0004-6361/201220616}

\bibitem[{{Crain} {et~al.}(2015){Crain}, {Schaye}, {Bower}, {Furlong},
  {Schaller}, {Theuns}, {Dalla Vecchia}, {Frenk}, {McCarthy}, {Helly},
  {Jenkins}, {Rosas-Guevara}, {White}, \& {Trayford}}]{Crain2015}
{Crain}, R.~A., {Schaye}, J., {Bower}, R.~G., {et~al.} 2015, \mnras, 450, 1937,
  \dodoi{10.1093/mnras/stv725}

\bibitem[{{Croom} {et~al.}(2012){Croom}, {Lawrence}, {Bland-Hawthorn},
  {Bryant}, {Fogarty}, {Richards}, {Goodwin}, {Farrell}, {Miziarski}, {Heald},
  {Jones}, {Lee}, {Colless}, {Brough}, {Hopkins}, {Bauer}, {Birchall}, {Ellis},
  {Horton}, {Leon-Saval}, {Lewis}, {L{\'o}pez-S{\'a}nchez}, {Min}, {Trinh}, \&
  {Trowland}}]{Croom2012}
{Croom}, S.~M., {Lawrence}, J.~S., {Bland-Hawthorn}, J., {et~al.} 2012, \mnras,
  421, 872, \dodoi{10.1111/j.1365-2966.2011.20365.x}

\bibitem[{{Curti} {et~al.}(2020){Curti}, {Mannucci}, {Cresci}, \&
  {Maiolino}}]{Curti2020}
{Curti}, M., {Mannucci}, F., {Cresci}, G., \& {Maiolino}, R. 2020, \mnras, 491,
  944, \dodoi{10.1093/mnras/stz2910}

\bibitem[{{Dopita} {et~al.}(2016){Dopita}, {Kewley}, {Sutherland}, \&
  {Nicholls}}]{Dopita2016}
{Dopita}, M.~A., {Kewley}, L.~J., {Sutherland}, R.~S., \& {Nicholls}, D.~C.
  2016, \apss, 361, 61, \dodoi{10.1007/s10509-016-2657-8}

\bibitem[{{Drory} {et~al.}(2015){Drory}, {MacDonald}, {Bershady}, {Bundy},
  {Gunn}, {Law}, {Smith}, {Stoll}, {Tremonti}, {Wake}, {Yan}, {Weijmans},
  {Byler}, {Cherinka}, {Cope}, {Eigenbrot}, {Harding}, {Holder}, {Huehnerhoff},
  {Jaehnig}, {Jansen}, {Klaene}, {Paat}, {Percival}, \& {Sayres}}]{Drory2015}
{Drory}, N., {MacDonald}, N., {Bershady}, M.~A., {et~al.} 2015, \aj, 149, 77,
  \dodoi{10.1088/0004-6256/149/2/77}

\bibitem[{{El-Badry} {et~al.}(2016){El-Badry}, {Wetzel}, {Geha}, {Hopkins},
  {Kere{\v{s}}}, {Chan}, \& {Faucher-Gigu{\`e}re}}]{Elbadry2016}
{El-Badry}, K., {Wetzel}, A., {Geha}, M., {et~al.} 2016, \apj, 820, 131,
  \dodoi{10.3847/0004-637X/820/2/131}

\bibitem[{{Ellison} {et~al.}(2008){Ellison}, {Patton}, {Simard}, \&
  {McConnachie}}]{Ellison2008}
{Ellison}, S.~L., {Patton}, D.~R., {Simard}, L., \& {McConnachie}, A.~W. 2008,
  \apjl, 672, L107, \dodoi{10.1086/527296}

\bibitem[{{Fang} {et~al.}(2013){Fang}, {Faber}, {Koo}, \& {Dekel}}]{Fang2013}
{Fang}, J.~J., {Faber}, S.~M., {Koo}, D.~C., \& {Dekel}, A. 2013, \apj, 776,
  63, \dodoi{10.1088/0004-637X/776/1/63}

\bibitem[{{Foreman-Mackey} {et~al.}(2013){Foreman-Mackey}, {Hogg}, {Lang}, \&
  {Goodman}}]{ForemanMackey2013}
{Foreman-Mackey}, D., {Hogg}, D.~W., {Lang}, D., \& {Goodman}, J. 2013, \pasp,
  125, 306, \dodoi{10.1086/670067}

\bibitem[{{Fraternali} \& {Binney}(2008)}]{Fraternali2008}
{Fraternali}, F., \& {Binney}, J.~J. 2008, \mnras, 386, 935,
  \dodoi{10.1111/j.1365-2966.2008.13071.x}

\bibitem[{{Gao} {et~al.}(2018){Gao}, {Wang}, {Kong}, {Lin}, {Liu}, {Liu},
  {Liu}, {Hu}, {Berhane Teklu}, {Chen}, \& {Zhao}}]{Gao2018}
{Gao}, Y., {Wang}, E., {Kong}, X., {et~al.} 2018, \apj, 868, 89,
  \dodoi{10.3847/1538-4357/aae9f1}

\bibitem[{{Garnett}(1990)}]{Garnett1990}
{Garnett}, D.~R. 1990, \apj, 363, 142, \dodoi{10.1086/169324}

\bibitem[{{Goddard} {et~al.}(2017){Goddard}, {Thomas}, {Maraston}, {Westfall},
  {Etherington}, {Riffel}, {Mallmann}, {Zheng}, {Argudo-Fern{\'a}ndez}, {Lian},
  {Bershady}, {Bundy}, {Drory}, {Law}, {Yan}, {Wake}, {Weijmans}, {Bizyaev},
  {Brownstein}, {Lane}, {Maiolino}, {Masters}, {Merrifield}, {Nitschelm},
  {Pan}, {Roman-Lopes}, {Storchi-Bergmann}, \& {Schneider}}]{Goddard2017}
{Goddard}, D., {Thomas}, D., {Maraston}, C., {et~al.} 2017, \mnras, 466, 4731,
  \dodoi{10.1093/mnras/stw3371}

\bibitem[{{Gunn} {et~al.}(2006){Gunn}, {Siegmund}, {Mannery}, {Owen}, {Hull},
  {Leger}, {Carey}, {Knapp}, {York}, {Boroski}, {Kent}, {Lupton}, {Rockosi},
  {Evans}, {Waddell}, {Anderson}, {Annis}, {Barentine}, {Bartoszek}, {Bastian},
  {Bracker}, {Brewington}, {Briegel}, {Brinkmann}, {Brown}, {Carr},
  {Czarapata}, {Drennan}, {Dombeck}, {Federwitz}, {Gillespie}, {Gonzales},
  {Hansen}, {Harvanek}, {Hayes}, {Jordan}, {Kinney}, {Klaene}, {Kleinman},
  {Kron}, {Kresinski}, {Lee}, {Limmongkol}, {Lindenmeyer}, {Long}, {Loomis},
  {McGehee}, {Mantsch}, {Neilsen}, {Neswold}, {Newman}, {Nitta}, {Peoples},
  {Pier}, {Prieto}, {Prosapio}, {Rivetta}, {Schneider}, {Snedden}, \&
  {Wang}}]{Gunn2006}
{Gunn}, J.~E., {Siegmund}, W.~A., {Mannery}, E.~J., {et~al.} 2006, \aj, 131,
  2332, \dodoi{10.1086/500975}

\bibitem[{Hunter(2007)}]{Matplotlib}
Hunter, J.~D. 2007, Computing in Science \& Engineering, 9, 90,
  \dodoi{10.1109/MCSE.2007.55}

\bibitem[{{Hwang} {et~al.}(2019){Hwang}, {Barrera-Ballesteros}, {Heckman},
  {Rowlands}, {Lin}, {Rodriguez-Gomez}, {Pan}, {Hsieh}, {S{\'a}nchez},
  {Bizyaev}, {S{\'a}nchez Almeida}, {Thilker}, {Lotz}, {Jones}, {Nair},
  {Andrews}, \& {Drory}}]{Hwang2019}
{Hwang}, H.-C., {Barrera-Ballesteros}, J.~K., {Heckman}, T.~M., {et~al.} 2019,
  \apj, 872, 144, \dodoi{10.3847/1538-4357/aaf7a3}

\bibitem[{{Kashino} {et~al.}(2016){Kashino}, {Renzini}, {Silverman}, \&
  {Daddi}}]{Kashino2016}
{Kashino}, D., {Renzini}, A., {Silverman}, J.~D., \& {Daddi}, E. 2016, \apjl,
  823, L24, \dodoi{10.3847/2041-8205/823/2/L24}

\bibitem[{{Kauffmann} {et~al.}(2003){Kauffmann}, {Heckman}, {Tremonti},
  {Brinchmann}, {Charlot}, {White}, {Ridgway}, {Brinkmann}, {Fukugita}, {Hall},
  {Ivezi{\'c}}, {Richards}, \& {Schneider}}]{Kauffmann2003}
{Kauffmann}, G., {Heckman}, T.~M., {Tremonti}, C., {et~al.} 2003, \mnras, 346,
  1055, \dodoi{10.1111/j.1365-2966.2003.07154.x}

\bibitem[{{Kennicutt}(1998)}]{Kennicutt1998}
{Kennicutt}, Robert~C., J. 1998, \apj, 498, 541, \dodoi{10.1086/305588}

\bibitem[{{Kewley} \& {Dopita}(2002)}]{Kewley2002}
{Kewley}, L.~J., \& {Dopita}, M.~A. 2002, \apjs, 142, 35,
  \dodoi{10.1086/341326}

\bibitem[{{Kewley} {et~al.}(2001){Kewley}, {Dopita}, {Sutherland}, {Heisler},
  \& {Trevena}}]{Kewley2001}
{Kewley}, L.~J., {Dopita}, M.~A., {Sutherland}, R.~S., {Heisler}, C.~A., \&
  {Trevena}, J. 2001, \apj, 556, 121, \dodoi{10.1086/321545}

\bibitem[{{Kewley} {et~al.}(2019){Kewley}, {Nicholls}, \&
  {Sutherland}}]{Kewley2019}
{Kewley}, L.~J., {Nicholls}, D.~C., \& {Sutherland}, R.~S. 2019, \araa, 57,
  511, \dodoi{10.1146/annurev-astro-081817-051832}

\bibitem[{{Kobulnicky} \& {Kewley}(2004)}]{Kobulnicky2004}
{Kobulnicky}, H.~A., \& {Kewley}, L.~J. 2004, \apj, 617, 240,
  \dodoi{10.1086/425299}

\bibitem[{Kolmogorov(1933)}]{Kolmogorov1933}
Kolmogorov, A. 1933, Inst. Ital. Attuari, Giorn., 4, 83.
\newblock \url{https://ci.nii.ac.jp/naid/10010480527/en/}

\bibitem[{{K{\"o}ppen} \& {Hensler}(2005)}]{Koppen2005}
{K{\"o}ppen}, J., \& {Hensler}, G. 2005, \aap, 434, 531,
  \dodoi{10.1051/0004-6361:20042266}

\bibitem[{{Kreckel} {et~al.}(2020){Kreckel}, {Ho}, {Blanc}, {Glover}, {Groves},
  {Rosolowsky}, {Bigiel}, {Boqu{\'\i}en}, {Chevance}, {Dale}, {Deger},
  {Emsellem}, {Grasha}, {Kim}, {Klessen}, {Kruijssen}, {Lee}, {Leroy}, {Liu},
  {McElroy}, {Meidt}, {Pessa}, {Sanchez-Blazquez}, {Sandstrom}, {Santoro},
  {Scheuermann}, {Schinnerer}, {Schruba}, {Utomo}, {Watkins}, \&
  {Williams}}]{Kreckel2020}
{Kreckel}, K., {Ho}, I.~T., {Blanc}, G.~A., {et~al.} 2020, \mnras, 499, 193,
  \dodoi{10.1093/mnras/staa2743}

\bibitem[{{Lacerda} {et~al.}(2018){Lacerda}, {Cid Fernandes}, {Couto},
  {Stasi{\'n}ska}, {Garc{\'\i}a-Benito}, {Vale Asari}, {P{\'e}rez},
  {Gonz{\'a}lez Delgado}, {S{\'a}nchez}, \& {de Amorim}}]{Lacerda2018}
{Lacerda}, E.~A.~D., {Cid Fernandes}, R., {Couto}, G.~S., {et~al.} 2018,
  \mnras, 474, 3727, \dodoi{10.1093/mnras/stx3022}

\bibitem[{{Law} {et~al.}(2015){Law}, {Yan}, {Bershady}, {Bundy}, {Cherinka},
  {Drory}, {MacDonald}, {S{\'a}nchez-Gallego}, {Wake}, {Weijmans}, {Blanton},
  {Klaene}, {Moran}, {Sanchez}, \& {Zhang}}]{Law2015}
{Law}, D.~R., {Yan}, R., {Bershady}, M.~A., {et~al.} 2015, \aj, 150, 19,
  \dodoi{10.1088/0004-6256/150/1/19}

\bibitem[{{Law} {et~al.}(2016){Law}, {Cherinka}, {Yan}, {Andrews}, {Bershady},
  {Bizyaev}, {Blanc}, {Blanton}, {Bolton}, {Brownstein}, {Bundy}, {Chen},
  {Drory}, {D'Souza}, {Fu}, {Jones}, {Kauffmann}, {MacDonald}, {Masters},
  {Newman}, {Parejko}, {S{\'a}nchez-Gallego}, {S{\'a}nchez}, {Schlegel},
  {Thomas}, {Wake}, {Weijmans}, {Westfall}, \& {Zhang}}]{Law2016}
{Law}, D.~R., {Cherinka}, B., {Yan}, R., {et~al.} 2016, \aj, 152, 83,
  \dodoi{10.3847/0004-6256/152/4/83}

\bibitem[{{Law} {et~al.}(2021){Law}, {Westfall}, {Bershady}, {Cappellari},
  {Yan}, {Belfiore}, {Bizyaev}, {Brownstein}, {Chen}, {Cherinka}, {Drory},
  {Lazarz}, \& {Shetty}}]{Law2021}
{Law}, D.~R., {Westfall}, K.~B., {Bershady}, M.~A., {et~al.} 2021, \aj, 161,
  52, \dodoi{10.3847/1538-3881/abcaa2}

\bibitem[{{Leitherer} {et~al.}(1999){Leitherer}, {Schaerer}, {Goldader},
  {Delgado}, {Robert}, {Kune}, {de Mello}, {Devost}, \&
  {Heckman}}]{Leitherer1999}
{Leitherer}, C., {Schaerer}, D., {Goldader}, J.~D., {et~al.} 1999, \apjs, 123,
  3, \dodoi{10.1086/313233}

\bibitem[{{Leroy} {et~al.}(2008){Leroy}, {Walter}, {Brinks}, {Bigiel}, {de
  Blok}, {Madore}, \& {Thornley}}]{Leroy2008}
{Leroy}, A.~K., {Walter}, F., {Brinks}, E., {et~al.} 2008, \aj, 136, 2782,
  \dodoi{10.1088/0004-6256/136/6/2782}

\bibitem[{{Li} {et~al.}(2021){Li}, {Krumholz}, {Wisnioski}, {Mendel}, {Kewley},
  {S{\'a}nchez}, \& {Galbany}}]{Li2021}
{Li}, Z., {Krumholz}, M.~R., {Wisnioski}, E., {et~al.} 2021, \mnras, 504, 5496,
  \dodoi{10.1093/mnras/stab1263}

\bibitem[{{Lian} {et~al.}(2019){Lian}, {Thomas}, {Li}, {Zheng}, {Maraston},
  {Bizyaev}, {Lane}, \& {Yan}}]{Lian2019}
{Lian}, J., {Thomas}, D., {Li}, C., {et~al.} 2019, \mnras, 489, 1436,
  \dodoi{10.1093/mnras/stz2218}

\bibitem[{{Licquia} \& {Newman}(2015)}]{Licquia2015}
{Licquia}, T.~C., \& {Newman}, J.~A. 2015, \apj, 806, 96,
  \dodoi{10.1088/0004-637X/806/1/96}

\bibitem[{{Lilly} {et~al.}(2013){Lilly}, {Carollo}, {Pipino}, {Renzini}, \&
  {Peng}}]{Lilly2013}
{Lilly}, S.~J., {Carollo}, C.~M., {Pipino}, A., {Renzini}, A., \& {Peng}, Y.
  2013, \apj, 772, 119, \dodoi{10.1088/0004-637X/772/2/119}

\bibitem[{{Luo} {et~al.}(2021){Luo}, {Heckman}, {Hwang}, {Rowlands},
  {S{\'a}nchez-Menguiano}, {Riffel}, {Bizyaev}, {Andrews},
  {Fern{\'a}ndez-Trincado}, {Drory}, {S{\'a}nchez Almeida}, {Maiolino}, {Lane},
  \& {Argudo-Fern{\'a}ndez}}]{Luo2021}
{Luo}, Y., {Heckman}, T., {Hwang}, H.-C., {et~al.} 2021, \apj, 908, 183,
  \dodoi{10.3847/1538-4357/abd1df}

\bibitem[{{Mannucci} {et~al.}(2010){Mannucci}, {Cresci}, {Maiolino}, {Marconi},
  \& {Gnerucci}}]{Mannucci2010}
{Mannucci}, F., {Cresci}, G., {Maiolino}, R., {Marconi}, A., \& {Gnerucci}, A.
  2010, \mnras, 408, 2115, \dodoi{10.1111/j.1365-2966.2010.17291.x}

\bibitem[{{Martig} {et~al.}(2009){Martig}, {Bournaud}, {Teyssier}, \&
  {Dekel}}]{Martig2009}
{Martig}, M., {Bournaud}, F., {Teyssier}, R., \& {Dekel}, A. 2009, \apj, 707,
  250, \dodoi{10.1088/0004-637X/707/1/250}

\bibitem[{{Masters} {et~al.}(2016){Masters}, {Faisst}, \&
  {Capak}}]{Masters2016}
{Masters}, D., {Faisst}, A., \& {Capak}, P. 2016, \apj, 828, 18,
  \dodoi{10.3847/0004-637X/828/1/18}

\bibitem[{{Matthee} \& {Schaye}(2018)}]{Matthee2018}
{Matthee}, J., \& {Schaye}, J. 2018, \mnras, 479, L34,
  \dodoi{10.1093/mnrasl/sly093}

\bibitem[{{McAlpine} {et~al.}(2016){McAlpine}, {Helly}, {Schaller}, {Trayford},
  {Qu}, {Furlong}, {Bower}, {Crain}, {Schaye}, {Theuns}, {Dalla Vecchia},
  {Frenk}, {McCarthy}, {Jenkins}, {Rosas-Guevara}, {White}, {Baes}, {Camps}, \&
  {Lemson}}]{McAlpine2016}
{McAlpine}, S., {Helly}, J.~C., {Schaller}, M., {et~al.} 2016, Astronomy and
  Computing, 15, 72, \dodoi{10.1016/j.ascom.2016.02.004}

\bibitem[{{Melioli} {et~al.}(2008){Melioli}, {Brighenti}, {D'Ercole}, \& {de
  Gouveia Dal Pino}}]{Melioli2008}
{Melioli}, C., {Brighenti}, F., {D'Ercole}, A., \& {de Gouveia Dal Pino}, E.~M.
  2008, \mnras, 388, 573, \dodoi{10.1111/j.1365-2966.2008.13446.x}

\bibitem[{{Melioli} {et~al.}(2009){Melioli}, {Brighenti}, {D'Ercole}, \& {de
  Gouveia Dal Pino}}]{Melioli2009}
---. 2009, \mnras, 399, 1089, \dodoi{10.1111/j.1365-2966.2009.14725.x}

\bibitem[{{Mingozzi} {et~al.}(2020){Mingozzi}, {Belfiore}, {Cresci}, {Bundy},
  {Bershady}, {Bizyaev}, {Blanc}, {Boquien}, {Drory}, {Fu}, {Maiolino},
  {Riffel}, {Schaefer}, {Storchi-Bergmann}, {Telles}, {Tremonti}, {Zakamska},
  \& {Zhang}}]{Mingozzi2020}
{Mingozzi}, M., {Belfiore}, F., {Cresci}, G., {et~al.} 2020, \aap, 636, A42,
  \dodoi{10.1051/0004-6361/201937203}

\bibitem[{{Moll{\'a}} {et~al.}(2006){Moll{\'a}}, {V{\'\i}lchez}, {Gavil{\'a}n},
  \& {D{\'\i}az}}]{Molla2006}
{Moll{\'a}}, M., {V{\'\i}lchez}, J.~M., {Gavil{\'a}n}, M., \& {D{\'\i}az},
  A.~I. 2006, \mnras, 372, 1069, \dodoi{10.1111/j.1365-2966.2006.10892.x}

\bibitem[{{Moran} {et~al.}(2012){Moran}, {Heckman}, {Kauffmann}, {Dav{\'e}},
  {Catinella}, {Brinchmann}, {Wang}, {Schiminovich}, {Saintonge},
  {Gracia-Carpio}, {Tacconi}, {Giovanelli}, {Haynes}, {Fabello}, {Hummels},
  {Lemonias}, \& {Wu}}]{Moran2012}
{Moran}, S.~M., {Heckman}, T.~M., {Kauffmann}, G., {et~al.} 2012, \apj, 745,
  66, \dodoi{10.1088/0004-637X/745/1/66}

\bibitem[{{Newville} {et~al.}(2014){Newville}, {Stensitzki}, {Allen}, \&
  {Ingargiola}}]{Newville2014}
{Newville}, M., {Stensitzki}, T., {Allen}, D.~B., \& {Ingargiola}, A. 2014,
  {LMFIT: Non-Linear Least-Square Minimization and Curve-Fitting for Python},
  0.8.0,  Zenodo, \dodoi{10.5281/zenodo.11813}

\bibitem[{{Nicholls} {et~al.}(2017){Nicholls}, {Sutherland}, {Dopita},
  {Kewley}, \& {Groves}}]{Nicholls2017}
{Nicholls}, D.~C., {Sutherland}, R.~S., {Dopita}, M.~A., {Kewley}, L.~J., \&
  {Groves}, B.~A. 2017, \mnras, 466, 4403, \dodoi{10.1093/mnras/stw3235}

\bibitem[{{O'Donnell}(1994)}]{ODonnell1994}
{O'Donnell}, J.~E. 1994, \apj, 422, 158, \dodoi{10.1086/173713}

\bibitem[{{Oppenheimer} {et~al.}(2010){Oppenheimer}, {Dav{\'e}}, {Kere{\v{s}}},
  {Fardal}, {Katz}, {Kollmeier}, \& {Weinberg}}]{Oppenheimer2010}
{Oppenheimer}, B.~D., {Dav{\'e}}, R., {Kere{\v{s}}}, D., {et~al.} 2010, \mnras,
  406, 2325, \dodoi{10.1111/j.1365-2966.2010.16872.x}

\bibitem[{{Pagel} {et~al.}(1992){Pagel}, {Simonson}, {Terlevich}, \&
  {Edmunds}}]{Pagel1992}
{Pagel}, B.~E.~J., {Simonson}, E.~A., {Terlevich}, R.~J., \& {Edmunds}, M.~G.
  1992, \mnras, 255, 325, \dodoi{10.1093/mnras/255.2.325}

\bibitem[{{Parikh} {et~al.}(2018){Parikh}, {Thomas}, {Maraston}, {Westfall},
  {Goddard}, {Lian}, {Meneses-Goytia}, {Jones}, {Vaughan}, {Andrews},
  {Bershady}, {Bizyaev}, {Brinkmann}, {Brownstein}, {Bundy}, {Drory},
  {Emsellem}, {Law}, {Newman}, {Roman-Lopes}, {Wake}, {Yan}, \&
  {Zheng}}]{Parikh2018}
{Parikh}, T., {Thomas}, D., {Maraston}, C., {et~al.} 2018, \mnras, 477, 3954,
  \dodoi{10.1093/mnras/sty785}

\bibitem[{{Peng} \& {Maiolino}(2014)}]{Peng2014}
{Peng}, Y.-j., \& {Maiolino}, R. 2014, \mnras, 438, 262,
  \dodoi{10.1093/mnras/stt2175}

\bibitem[{{P{\'e}rez-Montero} {et~al.}(2021){P{\'e}rez-Montero}, {Amor{\'\i}n},
  {S{\'a}nchez Almeida}, {V{\'\i}lchez}, {Garc{\'\i}a-Benito}, \&
  {Kehrig}}]{PerezMontero2021}
{P{\'e}rez-Montero}, E., {Amor{\'\i}n}, R., {S{\'a}nchez Almeida}, J., {et~al.}
  2021, \mnras, 504, 1237, \dodoi{10.1093/mnras/stab862}

\bibitem[{{P{\'e}rez-Montero} \& {Contini}(2009)}]{PerezMontero2009}
{P{\'e}rez-Montero}, E., \& {Contini}, T. 2009, \mnras, 398, 949,
  \dodoi{10.1111/j.1365-2966.2009.15145.x}

\bibitem[{{P{\'e}rez-Montero} {et~al.}(2013){P{\'e}rez-Montero}, {Contini},
  {Lamareille}, {Maier}, {Carollo}, {Kneib}, {Le F{\`e}vre}, {Lilly},
  {Mainieri}, {Renzini}, {Scodeggio}, {Zamorani}, {Bardelli}, {Bolzonella},
  {Bongiorno}, {Caputi}, {Cucciati}, {de la Torre}, {de Ravel}, {Franzetti},
  {Garilli}, {Iovino}, {Kampczyk}, {Knobel}, {Kova{\v{c}}}, {Le Borgne}, {Le
  Brun}, {Mignoli}, {Pell{\`o}}, {Peng}, {Presotto}, {Ricciardelli},
  {Silverman}, {Tanaka}, {Tasca}, {Tresse}, {Vergani}, \&
  {Zucca}}]{PerezMontero2013}
{P{\'e}rez-Montero}, E., {Contini}, T., {Lamareille}, F., {et~al.} 2013, \aap,
  549, A25, \dodoi{10.1051/0004-6361/201220070}

\bibitem[{{Pettini} \& {Pagel}(2004)}]{Pettini2004}
{Pettini}, M., \& {Pagel}, B.~E.~J. 2004, \mnras, 348, L59,
  \dodoi{10.1111/j.1365-2966.2004.07591.x}

\bibitem[{{Pilyugin} \& {Grebel}(2016)}]{Pilyugin2016}
{Pilyugin}, L.~S., \& {Grebel}, E.~K. 2016, \mnras, 457, 3678,
  \dodoi{10.1093/mnras/stw238}

\bibitem[{{Rosales-Ortega} {et~al.}(2012){Rosales-Ortega}, {S{\'a}nchez},
  {Iglesias-P{\'a}ramo}, {D{\'\i}az}, {V{\'\i}lchez}, {Bland-Hawthorn},
  {Husemann}, \& {Mast}}]{RosalesOrtega2012}
{Rosales-Ortega}, F.~F., {S{\'a}nchez}, S.~F., {Iglesias-P{\'a}ramo}, J.,
  {et~al.} 2012, \apjl, 756, L31, \dodoi{10.1088/2041-8205/756/2/L31}

\bibitem[{{Salpeter}(1955)}]{Salpeter1955}
{Salpeter}, E.~E. 1955, \apj, 121, 161, \dodoi{10.1086/145971}

\bibitem[{{S{\'a}nchez}(2020)}]{Sanchez2020ARAA}
{S{\'a}nchez}, S.~F. 2020, \araa, 58, 99,
  \dodoi{10.1146/annurev-astro-012120-013326}

\bibitem[{{S{\'a}nchez} {et~al.}(2021){S{\'a}nchez}, {Walcher},
  {Lopez-Cob{\'a}}, {Barrera-Ballesteros}, {Mej{\'\i}a-Narv{\'a}ez},
  {Espinosa-Ponce}, \& {Camps-Fari{\~n}a}}]{Sanchez2021}
{S{\'a}nchez}, S.~F., {Walcher}, C.~J., {Lopez-Cob{\'a}}, C., {et~al.} 2021,
  \rmxaa, 57, 3, \dodoi{10.22201/ia.01851101p.2021.57.01.01}

\bibitem[{{S{\'a}nchez} {et~al.}(2012){S{\'a}nchez}, {Rosales-Ortega},
  {Marino}, {Iglesias-P{\'a}ramo}, {V{\'\i}lchez}, {Kennicutt}, {D{\'\i}az},
  {Mast}, {Monreal-Ibero}, {Garc{\'\i}a-Benito}, {Bland -Hawthorn},
  {P{\'e}rez}, {Gonz{\'a}lez Delgado}, {Husemann}, {L{\'o}pez-S{\'a}nchez},
  {Cid Fernand es}, {Kehrig}, {Walcher}, {Gil de Paz}, \&
  {Ellis}}]{Sanchez2012}
{S{\'a}nchez}, S.~F., {Rosales-Ortega}, F.~F., {Marino}, R.~A., {et~al.} 2012,
  \aap, 546, A2, \dodoi{10.1051/0004-6361/201219578}

\bibitem[{{S{\'a}nchez} {et~al.}(2015){S{\'a}nchez}, {Galbany}, {P{\'e}rez},
  {S{\'a}nchez-Bl{\'a}zquez}, {Falc{\'o}n-Barroso}, {Rosales-Ortega},
  {S{\'a}nchez-Menguiano}, {Marino}, {Kuncarayakti}, {Anderson}, {Kruehler},
  {Cano-D{\'\i}az}, {Barrera-Ballesteros}, \&
  {Gonz{\'a}lez-Gonz{\'a}lez}}]{Sanchez2015}
{S{\'a}nchez}, S.~F., {Galbany}, L., {P{\'e}rez}, E., {et~al.} 2015, \aap, 573,
  A105, \dodoi{10.1051/0004-6361/201424950}

\bibitem[{{S{\'a}nchez} {et~al.}(2016{\natexlab{a}}){S{\'a}nchez}, {P{\'e}rez},
  {S{\'a}nchez-Bl{\'a}zquez}, {Gonz{\'a}lez}, {Ros{\'a}les-Ortega},
  {Cano-D{\'\i}az}, {L{\'o}pez-Cob{\'a}}, {Marino}, {Gil de Paz}, {Moll{\'a}},
  {L{\'o}pez-S{\'a}nchez}, {Ascasibar}, \&
  {Barrera-Ballesteros}}]{Sanchez2016a}
{S{\'a}nchez}, S.~F., {P{\'e}rez}, E., {S{\'a}nchez-Bl{\'a}zquez}, P., {et~al.}
  2016{\natexlab{a}}, \rmxaa, 52, 21.
\newblock \doarXiv{1509.08552}

\bibitem[{{S{\'a}nchez} {et~al.}(2016{\natexlab{b}}){S{\'a}nchez}, {P{\'e}rez},
  {S{\'a}nchez-Bl{\'a}zquez}, {Garc{\'\i}a-Benito}, {Ibarra-Mede},
  {Gonz{\'a}lez}, {Rosales-Ortega}, {S{\'a}nchez-Menguiano}, {Ascasibar},
  {Bitsakis}, {Law}, {Cano-D{\'\i}az}, {L{\'o}pez-Cob{\'a}}, {Marino}, {Gil de
  Paz}, {L{\'o}pez-S{\'a}nchez}, {Barrera-Ballesteros}, {Galbany}, {Mast},
  {Abril-Melgarejo}, \& {Roman-Lopes}}]{Sanchez2016b}
---. 2016{\natexlab{b}}, \rmxaa, 52, 171.
\newblock \doarXiv{1602.01830}

\bibitem[{{S{\'a}nchez} {et~al.}(2018){S{\'a}nchez}, {Avila-Reese},
  {Hernandez-Toledo}, {Cortes-Su{\'a}rez}, {Rodr{\'\i}guez-Puebla},
  {Ibarra-Medel}, {Cano-D{\'\i}az}, {Barrera-Ballesteros}, {Negrete},
  {Calette}, {de Lorenzo-C{\'a}ceres}, {Ortega-Minakata}, {Aquino},
  {Valenzuela}, {Clemente}, {Storchi-Bergmann}, {Riffel}, {Schimoia}, {Riffel},
  {Rembold}, {Brownstein}, {Pan}, {Yates}, {Mallmann}, \&
  {Bitsakis}}]{Sanchez2018}
{S{\'a}nchez}, S.~F., {Avila-Reese}, V., {Hernandez-Toledo}, H., {et~al.} 2018,
  \rmxaa, 54, 217.
\newblock \doarXiv{1709.05438}

\bibitem[{{Schaefer} {et~al.}(2020){Schaefer}, {Tremonti}, {Belfiore}, {Pace},
  {Bershady}, {Andrews}, \& {Drory}}]{Schaefer2020}
{Schaefer}, A.~L., {Tremonti}, C., {Belfiore}, F., {et~al.} 2020, \apjl, 890,
  L3, \dodoi{10.3847/2041-8213/ab6f06}

\bibitem[{{Schaefer} {et~al.}(2019){Schaefer}, {Tremonti}, {Pace}, {Belfiore},
  {Argudo-Fernandez}, {Bershady}, {Drory}, {Jones}, {Maiolino}, {Stark},
  {Wake}, \& {Yan}}]{Schaefer2019}
{Schaefer}, A.~L., {Tremonti}, C., {Pace}, Z., {et~al.} 2019, \apj, 884, 156,
  \dodoi{10.3847/1538-4357/ab43ca}

\bibitem[{{Schaye} {et~al.}(2015){Schaye}, {Crain}, {Bower}, {Furlong},
  {Schaller}, {Theuns}, {Dalla Vecchia}, {Frenk}, {McCarthy}, {Helly},
  {Jenkins}, {Rosas-Guevara}, {White}, {Baes}, {Booth}, {Camps}, {Navarro},
  {Qu}, {Rahmati}, {Sawala}, {Thomas}, \& {Trayford}}]{Schaye2015}
{Schaye}, J., {Crain}, R.~A., {Bower}, R.~G., {et~al.} 2015, \mnras, 446, 521,
  \dodoi{10.1093/mnras/stu2058}

\bibitem[{{Schmidt}(1959)}]{Schmidt1959}
{Schmidt}, M. 1959, \apj, 129, 243, \dodoi{10.1086/146614}

\bibitem[{{Shapiro} \& {Field}(1976)}]{Shapiro1976}
{Shapiro}, P.~R., \& {Field}, G.~B. 1976, \apj, 205, 762,
  \dodoi{10.1086/154332}

\bibitem[{{Smee} {et~al.}(2013){Smee}, {Gunn}, {Uomoto}, {Roe}, {Schlegel},
  {Rockosi}, {Carr}, {Leger}, {Dawson}, {Olmstead}, {Brinkmann}, {Owen},
  {Barkhouser}, {Honscheid}, {Harding}, {Long}, {Lupton}, {Loomis}, {Anderson},
  {Annis}, {Bernardi}, {Bhardwaj}, {Bizyaev}, {Bolton}, {Brewington}, {Briggs},
  {Burles}, {Burns}, {Castander}, {Connolly}, {Davenport}, {Ebelke}, {Epps},
  {Feldman}, {Friedman}, {Frieman}, {Heckman}, {Hull}, {Knapp}, {Lawrence},
  {Loveday}, {Mannery}, {Malanushenko}, {Malanushenko}, {Merrelli}, {Muna},
  {Newman}, {Nichol}, {Oravetz}, {Pan}, {Pope}, {Ricketts}, {Shelden},
  {Sandford}, {Siegmund}, {Simmons}, {Smith}, {Snedden}, {Schneider},
  {SubbaRao}, {Tremonti}, {Waddell}, \& {York}}]{Smee2013}
{Smee}, S.~A., {Gunn}, J.~E., {Uomoto}, A., {et~al.} 2013, \aj, 146, 32,
  \dodoi{10.1088/0004-6256/146/2/32}

\bibitem[{{Speagle} {et~al.}(2014){Speagle}, {Steinhardt}, {Capak}, \&
  {Silverman}}]{Speagle2014}
{Speagle}, J.~S., {Steinhardt}, C.~L., {Capak}, P.~L., \& {Silverman}, J.~D.
  2014, \apjs, 214, 15, \dodoi{10.1088/0067-0049/214/2/15}

\bibitem[{{Spitoni} {et~al.}(2009){Spitoni}, {Matteucci}, {Recchi}, {Cescutti},
  \& {Pipino}}]{Spitoni2009}
{Spitoni}, E., {Matteucci}, F., {Recchi}, S., {Cescutti}, G., \& {Pipino}, A.
  2009, \aap, 504, 87, \dodoi{10.1051/0004-6361/200911768}

\bibitem[{{Teklu} {et~al.}(2020){Teklu}, {Gao}, {Kong}, {Lin}, \&
  {Liang}}]{Teklu2020}
{Teklu}, B.~B., {Gao}, Y., {Kong}, X., {Lin}, Z., \& {Liang}, Z. 2020, \apj,
  897, 61, \dodoi{10.3847/1538-4357/ab94af}

\bibitem[{{Thurston} {et~al.}(1996){Thurston}, {Edmunds}, \&
  {Henry}}]{Thurston1996}
{Thurston}, T.~R., {Edmunds}, M.~G., \& {Henry}, R.~B.~C. 1996, \mnras, 283,
  990, \dodoi{10.1093/mnras/283.3..990}

\bibitem[{{Tinsley}(1980)}]{Tinsley1980}
{Tinsley}, B.~M. 1980, \fcp, 5, 287

\bibitem[{{Tremonti} {et~al.}(2004){Tremonti}, {Heckman}, {Kauffmann},
  {Brinchmann}, {Charlot}, {White}, {Seibert}, {Peng}, {Schlegel}, {Uomoto},
  {Fukugita}, \& {Brinkmann}}]{Tremonti2004}
{Tremonti}, C.~A., {Heckman}, T.~M., {Kauffmann}, G., {et~al.} 2004, \apj, 613,
  898, \dodoi{10.1086/423264}

\bibitem[{{Vale Asari} {et~al.}(2019){Vale Asari}, {Couto}, {Cid Fernandes},
  {Stasi{\'n}ska}, {de Amorim}, {Ruschel-Dutra}, {Werle}, \&
  {Florido}}]{ValeAsari2019}
{Vale Asari}, N., {Couto}, G.~S., {Cid Fernandes}, R., {et~al.} 2019, \mnras,
  489, 4721, \dodoi{10.1093/mnras/stz2470}

\bibitem[{{Vincenzo} {et~al.}(2016){Vincenzo}, {Belfiore}, {Maiolino},
  {Matteucci}, \& {Ventura}}]{Vincenzo2016}
{Vincenzo}, F., {Belfiore}, F., {Maiolino}, R., {Matteucci}, F., \& {Ventura},
  P. 2016, \mnras, 458, 3466, \dodoi{10.1093/mnras/stw532}

\bibitem[{{Wake} {et~al.}(2017){Wake}, {Bundy}, {Diamond-Stanic}, {Yan},
  {Blanton}, {Bershady}, {S{\'a}nchez-Gallego}, {Drory}, {Jones}, \&
  {Kauffmann}}]{Wake2017}
{Wake}, D.~A., {Bundy}, K., {Diamond-Stanic}, A.~M., {et~al.} 2017, \aj, 154,
  86, \dodoi{10.3847/1538-3881/aa7ecc}

\bibitem[{{Wang} \& {Lilly}(2020)}]{Wang2020}
{Wang}, E., \& {Lilly}, S.~J. 2020, arXiv e-prints, arXiv:2009.01935.
\newblock \doarXiv{2009.01935}

\bibitem[{{Westfall} {et~al.}(2019){Westfall}, {Cappellari}, {Bershady},
  {Bundy}, {Belfiore}, {Ji}, {Law}, {Schaefer}, {Shetty}, {Tremonti}, {Yan},
  {Andrews}, {Brownstein}, {Cherinka}, {Coccato}, {Drory}, {Maraston},
  {Parikh}, {S{\'a}nchez-Gallego}, {Thomas}, {Weijmans}, {Barrera-Ballesteros},
  {Du}, {Goddard}, {Li}, {Masters}, {Ibarra Medel}, {S{\'a}nchez}, {Yang},
  {Zheng}, \& {Zhou}}]{Westfall2019}
{Westfall}, K.~B., {Cappellari}, M., {Bershady}, M.~A., {et~al.} 2019, arXiv
  e-prints.
\newblock \doarXiv{1901.00856}

\bibitem[{{Wilkinson} {et~al.}(2017){Wilkinson}, {Maraston}, {Goddard},
  {Thomas}, \& {Parikh}}]{Wilkinson2017}
{Wilkinson}, D.~M., {Maraston}, C., {Goddard}, D., {Thomas}, D., \& {Parikh},
  T. 2017, \mnras, 472, 4297, \dodoi{10.1093/mnras/stx2215}

\bibitem[{{Worthey} \& {Ottaviani}(1997)}]{Worthey1997}
{Worthey}, G., \& {Ottaviani}, D.~L. 1997, \apjs, 111, 377,
  \dodoi{10.1086/313021}

\bibitem[{{Yan} {et~al.}(2016{\natexlab{a}}){Yan}, {Tremonti}, {Bershady},
  {Law}, {Schlegel}, {Bundy}, {Drory}, {MacDonald}, {Bizyaev}, {Blanc},
  {Blanton}, {Cherinka}, {Eigenbrot}, {Gunn}, {Harding}, {Hogg},
  {S{\'a}nchez-Gallego}, {S{\'a}nchez}, {Wake}, {Weijmans}, {Xiao}, \&
  {Zhang}}]{Yan2016b}
{Yan}, R., {Tremonti}, C., {Bershady}, M.~A., {et~al.} 2016{\natexlab{a}}, \aj,
  151, 8, \dodoi{10.3847/0004-6256/151/1/8}

\bibitem[{{Yan} {et~al.}(2016{\natexlab{b}}){Yan}, {Bundy}, {Law}, {Bershady},
  {Andrews}, {Cherinka}, {Diamond-Stanic}, {Drory}, {MacDonald},
  {S{\'a}nchez-Gallego}, {Thomas}, {Wake}, {Weijmans}, {Westfall}, {Zhang},
  {Arag{\'o}n-Salamanca}, {Belfiore}, {Bizyaev}, {Blanc}, {Blanton},
  {Brownstein}, {Cappellari}, {D'Souza}, {Emsellem}, {Fu}, {Gaulme}, {Graham},
  {Goddard}, {Gunn}, {Harding}, {Jones}, {Kinemuchi}, {Li}, {Li}, {Maiolino},
  {Mao}, {Maraston}, {Masters}, {Merrifield}, {Oravetz}, {Pan}, {Parejko},
  {Sanchez}, {Schlegel}, {Simmons}, {Thanjavur}, {Tinker}, {Tremonti}, {van den
  Bosch}, \& {Zheng}}]{Yan2016}
{Yan}, R., {Bundy}, K., {Law}, D.~R., {et~al.} 2016{\natexlab{b}}, \aj, 152,
  197, \dodoi{10.3847/0004-6256/152/6/197}

\bibitem[{{Yan} {et~al.}(2019){Yan}, {Chen}, {Lazarz}, {Bizyaev}, {Maraston},
  {Stringfellow}, {McCarthy}, {Meneses-Goytia}, {Law}, {Thomas}, {Falcon
  Barroso}, {S{\'a}nchez-Gallego}, {Schlafly}, {Zheng}, {Argudo-Fern{\'a}ndez},
  {Beaton}, {Beers}, {Bershady}, {Blanton}, {Brownstein}, {Bundy}, {Chambers},
  {Cherinka}, {De Lee}, {Drory}, {Galbany}, {Holtzman}, {Imig}, {Kaiser},
  {Kinemuchi}, {Liu}, {Luo}, {Magnier}, {Majewski}, {Nair}, {Oravetz},
  {Oravetz}, {Pan}, {Sobeck}, {Stassun}, {Talbot}, {Tremonti}, {Waters},
  {Weijmans}, {Wilhelm}, {Zasowski}, {Zhao}, \& {Zhao}}]{Yan2019}
{Yan}, R., {Chen}, Y., {Lazarz}, D., {et~al.} 2019, \apj, 883, 175,
  \dodoi{10.3847/1538-4357/ab3ebc}

\end{thebibliography}


\label{lastpage}

\end{document}